\documentclass[usenatbib]{mnras}

\usepackage{mathptmx}

\usepackage[T1]{fontenc}
\usepackage{ae,aecompl}
\usepackage{multicol}        

\usepackage{graphicx}   
\usepackage{amsmath}    
\usepackage{amssymb}    
\usepackage{threeparttable}
\usepackage{times}             
\usepackage{epstopdf}
\usepackage{natbib}
\usepackage{ulem}

\title[Chemical abundances of 20 barium stars]
{Chemical abundances of 20 barium stars from the OHP spectra}

\author[G. C. Yang et al.]{
Guochao Yang$^{1,2}$,
Jingkun Zhao$^{2}$\thanks{E-mail: zjk@nao.cas.cn},
Yanchun Liang$^{2}$\thanks{E-mail: ycliang@nao.cas.cn},
Monique Spite$^{3}$,
Francois Spite$^{3}$,
Jianrong Shi$^{2,4}$,
\newauthor
Shuai Liu$^{2}$,
Nian Liu$^{1}$,
Wenyuan Cui$^{5}$
and Gang Zhao$^{2}$
\\
$^{1}$School of Physics and Astronomy, China West Normal University, 637002 Nanchong, PR China\\
$^{2}$Key Laboratory of Optical Astronomy, National Astronomical Observatories, Chinese Academy of Sciences, 100012 Beijing, PR China\\
$^{3}$GEPI, Observatoire de Paris-Meudon, 92195 Meudon, France\\
$^{4}$School of Astronomy and Space Science, University of Chinese Academy of Sciences, 100049 Beijing, PR China\\
$^{5}$Department of Physics, Hebei Normal University, 050024 Shijiazhuang, PR China\\
}

\date{Accepted 2024 September 29. Received 2024 September 29; in original form 2024 April 18}

\pubyear{XXX}

\begin{document}
\label{firstpage}
\pagerange{\pageref{firstpage}--\pageref{lastpage}}
\maketitle

\begin{abstract}
Based on the high resolution and high signal-to-noise spectra, we derived the chemical abundances of 20 elements for
20 barium (Ba-) stars. For the first time, the detailed abundances of four sample stars, namely HD\,92482, HD\,150430,
HD\,151101 and HD\,177304 have been analyzed. Additionally, Ba element abundance has been measured using high resolution
spectra for the first time in six of the other 16 sample stars.
Based on the [s/Fe] ratios, the Ba-unknown star HD\,115927 can be classified as
a strong Ba-star, while the Ba-likely star HD\,160538 can be categorized into a mild Ba-star. Consequently, our sample
comprises three strong and 17 mild Ba-stars. The light odd-Z metal elements and Fe-peak elements exhibit near-solar
abundances. The [$\alpha$/Fe] ratios demonstrate decreasing trends with increasing metallicity. Moreover, the abundances
of n-capture elements show significant enhancements in different degrees. Using a threshold of the signed distances to
the solar r-process abundance pattern $d_{s}$ = 0.6, we find that all of our sample stars are normal Ba-stars, indicating that the
enhancements of s-process elements should be attributed to material transfer from their companions. We compare the observed
n-capture patterns of sample stars with the FRUITY models, and estimate the mass of the Thermally-Pulsing Asymptotic Giant
Branch stars that previously contaminated the Ba-stars. The models with low masses can successfully explain the observations.
From a kinematic point of view, we note that most of our sample stars are linked with the thin disk, while HD\,130255 may
be associated with the thick disk.
\end{abstract}




\section{Introduction}

Various nucleosynthesis processes operating in different mass stars are responsible for
generating diverse chemical elements \citep{Sneden2008}. The light metal elements from C to Zn can be
produced through charged-particle fusion reactions, whereas the elements heavier than
Fe-peak are predominantly synthesized by neutron-capture (n-capture) process \citep{Burbidge1957,
Busso2001,Ji2018,Cowan2021}. Based on the timescales of n-capture and radioactive decay
of unstable nuclei, the n-capture nucleosynthesis is generally categorized into slow (s-)
process, primarily occurring in asymptotic giant branch (AGB) stars and rapid (r-) process,
commonly associated with violent astrophysical environments such as
Type\,II supernovae (SNe\,II) explosions or neutron star mergers (NSMs)
\citep{Qian2007,Abbott2017,Watson2019,Stancliffe2021}. Abundance analysis of chemically
peculiar stars can help us investigating the nucleosynthesis processes of elements.

Barium (Ba-) stars are a significant class which shows anomalously strong lines of s-process
elements (e.g., Sr, Y, Ba and La) in their spectra and were first defined by \citet{Bidelman1951}.
Ba-stars are giants or dwarfs with spectral-types from $\sim$ G to K and effective temperatures
from $\sim$ 4000 to 6000\,K. The variations of radial velocity, together with the enhancement
of s-process elements, imply that the observed Ba-star should have an unseen white dwarf (WD)
companion \citep{McClure1980}. In other words, maybe all of the Ba-stars are in binary systems,
where the Ba-stars obtained s-process material from the AGB companions, which now evolved into WDs
\citep{McClure1983,Jorissen1998,Jorissen2019}. Furthermore, some Ba-stars (e.g., HD\,48565
and HD\,114520) are found in triple systems \citep{Norris2000,Escorza2019}.

The chemical abundance features of Ba-stars can provide important information for understanding
the different nucleosynthesis processes and the mass accretion in binaries or triples
\citep{Cseh2018,Stancliffe2021}. Based on the strength of \ion{Ba}{ii} line at $\lambda$ = 4554\,\AA~
and the scale of Ba intensities proposed by \citet{Warner1965}, \citet{Lu1991} compiled a catalogue
containing 389 Ba-star candidates. In the past decades, numerous studies have analyzed chemical abundances of
Ba-stars using high resolution and high signal-to-noise (S/N) spectra \cite[e.g.,][]{Zacs1994,Liang2000,
Boyarchuk2002,Liang2003,Antipova2004,Yushchenko2004,Allen2006a,
Allen2006b,Smiljanic2007,Pereira2009,Liu2009,Pereira2011,Katime2013}.

In recent years, \citet{deCastro2016} conducted a study on a large sample of 182 Ba-stars and determined
the abundances of Na, Al, $\alpha$-elements, Fe-peak elements, and the s-process elements
Y, Zr, La, Ce and Nd. They found that the [s/Fe] and [hs/ls] (where hs refers to the heavier
s-process elements, such as Ba, La, Ce and Nd, and ls for the lighter s-process elements, such as
Sr, Y and Zr) show strong anticorrelations with metallicity \cite[see also][]{Goriely2000,Busso2001}.
Based on the data
set of chemical abundances derived by \citet{deCastro2016}, \citet{Cseh2018} compared [hs/ls] ratios of
169 Ba-stars to those from the theoretical AGB models, and found a good consistency between the
observed [hs/ls] ratios and theoretical low-mass ($\lesssim$ 3-4\,$M_{\odot}$) AGB models, where $^{13}$C
is the main neutron source. \citet{Kong2018a} studied three Ba-dwarfs, and provided stellar mass and
abundances of light metal, $\alpha$-, Fe-peak and s-process elements. They suggested that the formation
scenarios of s-process enhancement in Ba-dwarfs are similar to those of Ba-giants, where the s-process
elements originate from the WD companions during the AGB phase through Roche-lobe overflow or wind accretion.

Analyzing chemical abundances of 18 Ba-stars using high resolution and high S/N ratio spectra from Keck/HIRES,
\citet{Liu2021} discovered two new Ba-giants (HD\,16178 and HD\,22233) and one Ba-subgiant (HD\,2946).
\citet{Roriz2021a} and \citet{Roriz2021b} investigated a large sample of 180 Ba-stars, presenting Rb and
n-capture elements, e.g., Sr, Nb, Mo, Ru, La, Sm and Eu, and compared the observed [hs/ls] and [Eu/La] ratios with
those from s-process models. They found that the [Nb,Mo,Ru/Sr] and [Ce,Nd,Sm/La] ratios of most sample
stars are higher than those of the model predictions, suggesting the need of new theoretical and
observational work to explain these features.
Considering the mass transfer in Ba binaries, \citet{Cseh2022} compared observed abundance patterns of
28 Ba-stars to the AGB models of Monash \citep{Lugaro2012,Fishlock2014,Karakas2016,Karakas2018} and
FRUITY \citep{Cristallo2009,Cristallo2011,Cristallo2015,Cristallo2016}. They concluded that the abundance
patterns of n-capture elements in most of the Ba-stars are well represented by the ejecta of low-mass non-rotating AGB
stars, where $^{13}$C is the main source of neutrons. Additionally, they found that some Ba-stars represent
abundance features of not only s-process but also other nucleosynthetic processes.
By utilizing FRUITY models, \citet{Goswami2023} conducted mass estimations for the primary companions of 158
Ba-stars, comprising 52 mild and 106 strong Ba-stars. Their analysis revealed that the masses of
primary companions in strong Ba-stars reach a peak at 2.5\,$M_{\odot}$ with a standard deviation of
0.51\,$M_{\odot}$, whereas those in mild Ba-stars peak at 3.7\,$M_{\odot}$ with a standard deviation of
1.03\,$M_{\odot}$. Through a comparison of the observed n-capture patterns of four Ba-stars with Monash and
FRUITY nucleosynthesis models, \citet{Roriz2024} determined that the low-mass AGB stars, responsible for the
pollution, effectively replicate the observed data.

\citet{Han1995} estimated that there may exist 800-22000 Ba-stars with magnitude V $\leq$ 10.0
in our Galaxy. The anomalies in elemental abundances of Ba-stars reflect the nature of the nucleosynthetic
processes occurring within the interior of AGB stars. Given that the different nucleosynthetic processes
inside various AGB stars are not yet fully understood, studying more Ba-stars is of particular interest.
To explore the abundance characteristics of Ba-stars in detail, we obtained high resolution and high S/N
spectra for 31 Ba-stars using the 2.16\,m telescope at Xinglong station of the National Astronomical
Observatories, Chinese Academy of Sciences (NAOC), and 20 Ba-stars using the 1.93\,m telescope at the
Haute-Provence Observatory (OHP, France) during 2001 and 2005. The chemical abundances of the former 31
Ba-stars have been analyzed in our previous papers \cite[see details in][]{Liang2003,Liu2009,Yang2016}.

In this work, we observed the spectra of the latter 20 Ba-stars at OHP and derived the abundances
of light metal, $\alpha$-, Fe-peak and n-capture elements.
In Sect.\,2, we present the observations and data reductions of the sample stars.
The stellar atmospheric parameters are provided in Sect.\,3. The results are shown in Sect.\,4.
Discussions are presented in Sect.\,5. In Sect.\,6, we conduct a kinematic investigation of
the sample stars. The conclusions are drawn in Sect.\,7.

\section{Observations and data reduction}

\begin{figure}
\centering
\includegraphics[width=\columnwidth]{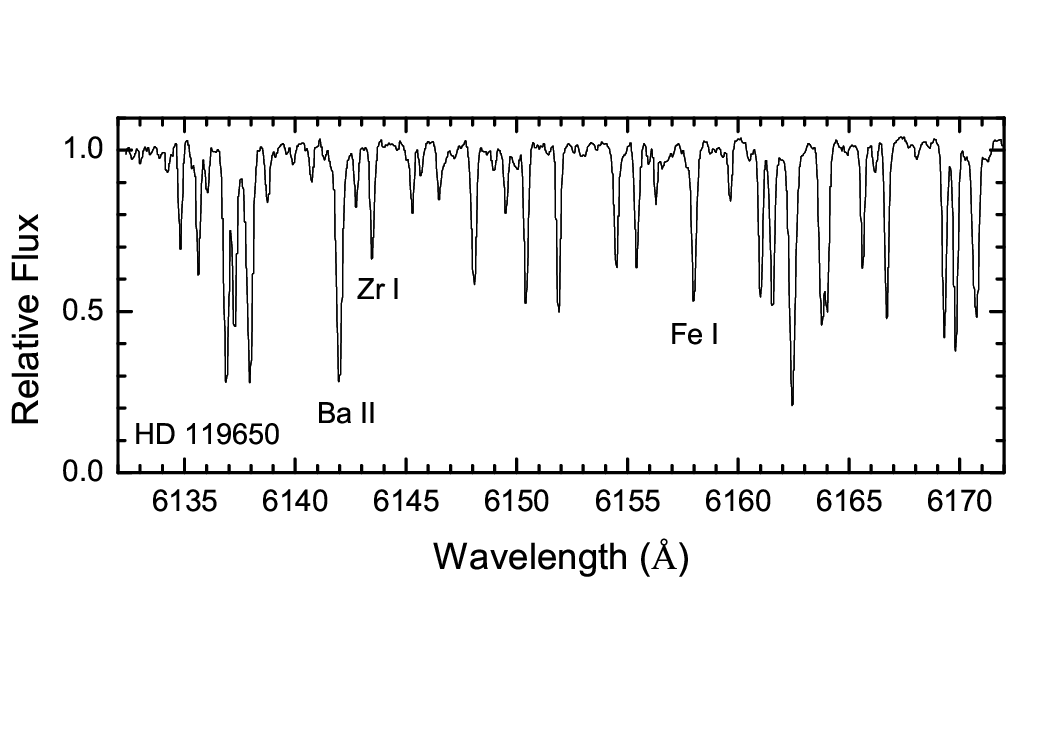}
\vspace{-1.5cm}
\caption{A portion of spectrum of the sample star HD\,119650.}
\label{fig:sample}
\end{figure}

The sample stars investigated in this paper were selected from \citet{Lu1991} and \citet{Jorissen1998}.
The basic parameters are listed in Table\,\ref{tab:basic}. The spectra of the sample stars were obtained
using the ELODIE spectrograph installed on the 1.93\,m telescope at OHP. The high resolution and high S/N
ratio spectra were acquired during a single observing run in 2003 (see details in Table\,\ref{tab:basic}).
The spectra cover a wavelength range of 4000 $\sim$ 7500\,\AA~with a resolving power of R $\sim$ 48000.
For most of the sample stars, the S/N exceeds 100 in the entire wavelength region. A portion of the
spectrum of the sample star HD\,119650 is shown in Fig.\,\ref{fig:sample}.

\begin{table*}
\begin{center}
\caption[]{Log of observations and basic data of the sample stars.}
\label{tab:basic}
\begin{threeparttable}
\begin{tabular}{llcllcrc}
  \hline\noalign{\smallskip}
Star name & Date & Exp. time (sec) & Ba & Sp. & $V$ & $K$ & Source \\
  \hline\noalign{\smallskip}
HD\,77912 & 2003 May 23 & 300 & 0.2m & G7II & 4.54 & 2.40 & 1 \\
HD\,85503 & 2003 May 23 & 100 & 1.0: & K2III & 3.88 & 1.22 & 1 \\
HD\,92482 & 2003 May 23 & 3600 & 3.0 & G8III & 8.62 & 6.32 & 1 \\
HD\,119650 & 2003 May 23 & 1800 & 0.3m & K1II & 7.59 & 4.91 & 1 \\
HD\,125079 & 2003 May 23 & 3600 & 2.0m & G2III & 8.63 & 6.61 & 1 \\
HD\,150430 & 2003 May 23 & 3500 & 0.5: & K0 & 8.19 & 5.88 & 1 \\
HD\,160507 & 2003 May 23 & 1100 & 1.0: & G5III & 6.56 & 4.39 & 1 \\
HD\,160538 & 2003 May 23 & 1200 & bl    & K0 & 6.63 & 4.09 & 2 \\
HD\,177304 & 2003 May 23 & 3600 & 0.5: & K & 8.61 & 5.81 & 1 \\
HD\,107950 & 2003 May 24 & 920 & 0.2 & G7III & 4.76 & 2.82 & 1 \\
HD\,133208 & 2003 May 24 & 300 & 0.5 & G8III & 3.52 & 1.34 & 1 \\
HD\,115927 & 2003 May 24 & 3600 & ? & K0II & 7.70 & 5.18 & 1 \\
HD\,130255 & 2003 May 24 & 3600 & 1.0c & G9V & 8.64 & 5.83 & 2 \\
HD\,168532 & 2003 May 24 & 900 & 0.5 & K3III & 5.28 & 1.85 & 1 \\
HD\,151101 & 2003 May 24 & 600 & 0.3 & K0III & 4.84 & 2.11 & 1 \\
HD\,176524 & 2003 May 24 & 600 & 0.2 & K0III & 4.81 & 2.32 & 1 \\
HD\,176670 & 2003 May 24 & 600 & 0.5 & K2.5III & 4.93 & 1.57 & 1 \\
HD\,101079 & 2003 May 27 & 3600 & 1.0 & K1 & 8.18 & 6.02 & 2 \\
HD\,131873 & 2003 May 27 & 120 & 0.3 & K4III & 2.08 & $-$1.39 & 1 \\
HD\,136138 & 2003 May 27 & 900 & 0.3 & G8III & 5.69 & 3.61 & 1 \\
  \noalign{\smallskip}\hline\\
\end{tabular}
\begin{tablenotes}
\item Notes. Ba intensity in Col. 4. c: certain Ba-stars, m: marginal Ba-stars,
bl: Ba-likely stars, `:' refers to the case of Ba intensity being uncertain, and
`?' means the Ba intensity being unidentified.
\item References: 1. \citet{Lu1991}; 2. \citet{Jorissen1998}.
\end{tablenotes}
\end{threeparttable}
\end{center}
\end{table*}
The 1D wavelength-calibrated spectra were retrieved from
the ELODIE\footnote{http://atlas.obs-hp.fr/elodie/} archive \citep{Moultaka2004}.
We normalized the spectra by fitting the continuum using cubic spline functions
with smooth parameters and computed stellar radial velocities by fitting
the profiles of approximately 80 pre-selected lines \citep{Wang2011}.
The measurements of equivalent width (EW) are performed using two methods: Gaussian function
fitting and direct integration. A Gaussian fitting is adopted for weak lines, while direct integration is
used for unblended strong lines \citep{Zhao2000}. Lines with 20\,m\AA\,$<$ EW $<$ 120\,m\AA~have been
chosen for the following abundance calculation of Fe. For other elements, the lines with
20\,m\AA\,$<$ EW $<$ 250\,m\AA~have been adopted due to limited availability. Furthermore,
close attention is paid to ensuring the consistency of the continuum level.

\begin{figure}
\centering
\includegraphics[width=\columnwidth]{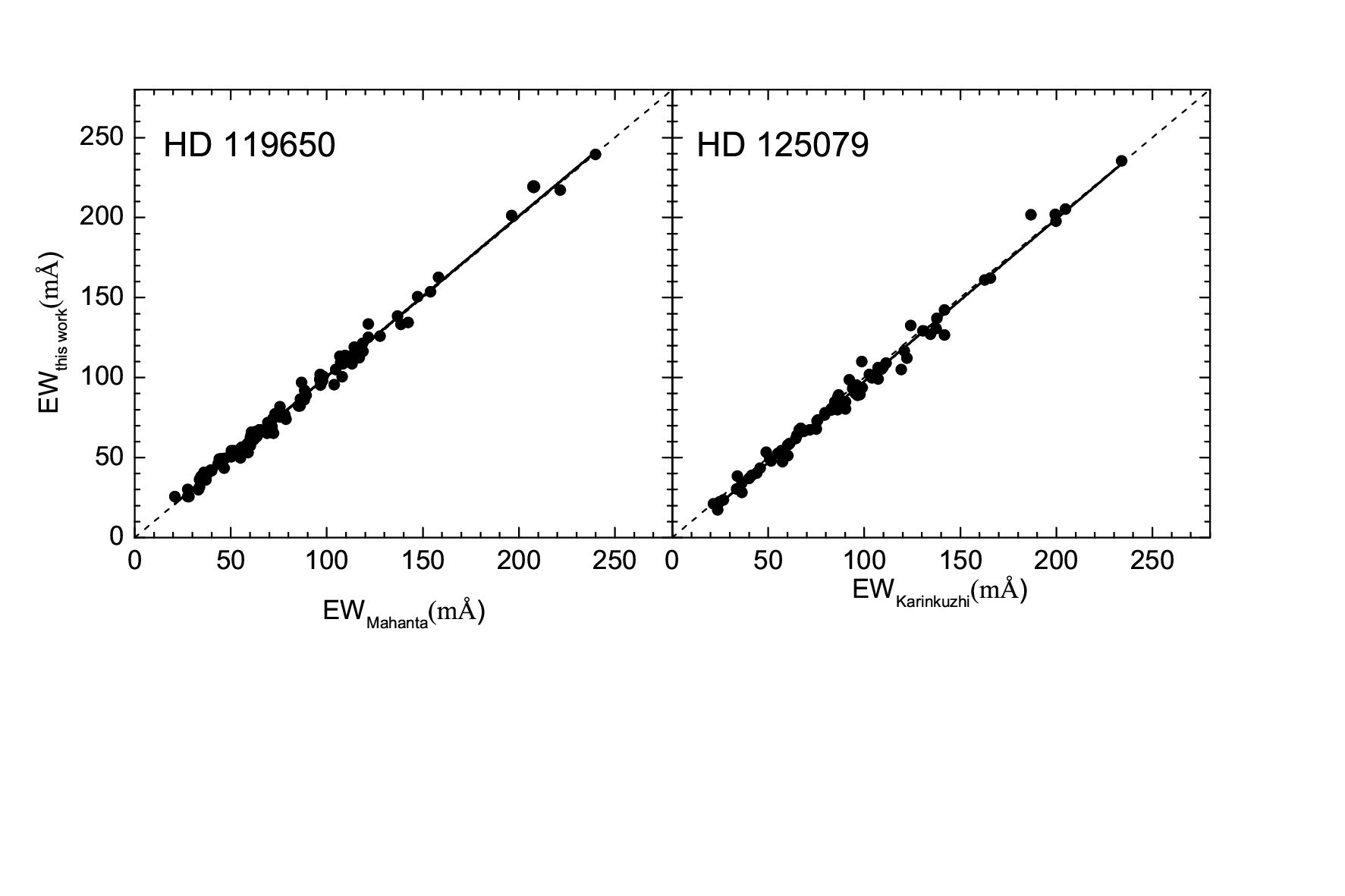}
\vspace{-2cm}
\caption{Comparisons of EWs of common lines measured in this work with those of \citet{Mahanta2016}
and \citet{Karinkuzhi2014} for the sample star HD\,119650 (the left panel) and HD\,125079 (the right panel),
respectively. The solid lines are the linear fits to the measured points. The dashed lines are
one-to-one correlations.}
\label{fig:ew}
\end{figure}

For assessing the reliability of the EW measurements in this study, we take HD\,119650
and HD\,125079 as sample stars and compared the EWs of common lines measured in our
work with those reported by \citet{Mahanta2016} and \citet{Karinkuzhi2014}.
The comparison results are shown in Fig.\,\ref{fig:ew}, revealing a mean deviation of about 3.5\,m\AA~for
the EWs of the common lines between our study and the previous investigations.

\section{Stellar atmospheric parameters}

The determination of the stellar atmospheric parameters, i.e., the effective temperature $T_{\rm eff}$,
surface gravity $\log$\,$g$, metallicity [Fe/H] and micro-turbulent velocity $\xi_{\rm t}$, is essential
for establishing elemental abundances. The effective temperatures were determined using two methods. Initially,
the effective temperatures were obtained from the photometric color $V-K$ utilizing the calibration provided
by \citet{Alonso1999}. The $V-K$ color data were taken from the SIMBAD\footnote{http://simbad.u-strasbg.fr/Simbad}
database, and the color excess $E$($V-K$) was computed from the 3D dust maps of \citet{Green2019}. The derived
effective temperatures and corresponding uncertainties are listed in Table\,\ref{tab:parameter} (Columns\,2 and 3).
The large uncertainties in the effective temperatures of HD\,77912, HD\,107950, HD\,136138, HD\,151101, HD\,160538
and HD\,176524 can be attributed to the large uncertainties in their $V-K$ values ($\gtrsim$ 0.30\,dex).
The effective temperatures can also be spectroscopically determined by ensuring that the iron abundances derived from
\ion{Fe}{i} lines are independent of the excitation potentials, and we get the final results when the slope of
[\ion{Fe}{i}/H] versus excitation potential is less than 0.005. We estimated the uncertainties from this on temperature value,
and the derived results are presented in Columns 4 and 5 of Table\,\ref{tab:parameter}. In our analysis, the spectroscopic
temperatures are derived from a vast array of clean Fe lines to calculate the elemental abundances of the sample stars.
\begin{table*}
\begin{center}
\caption[]{Physical parameters of the sample stars.}
\label{tab:parameter}
{\fontsize{7}{10}\selectfont
\begin{tabular}{lcrccccccrcrcc}
  \hline\noalign{\smallskip}
Star name & $T_{\rm eff}$ (K) & $\Delta {T_{\rm eff}}$ (K) & $T_{\rm eff}$ (K) & $\Delta {T_{\rm eff}}$ (K)
& $\log$\,$g$ & $\Delta {\log g}$ & $\log$\,$g$ & $\Delta {\log g}$ & [\ion{Fe}{i}/H] & $\Delta {\rm [Fe/H]}$
& [\ion{Fe}{ii}/H] & $\xi_{\rm t}$ (km\,s$^{-1}$) & $\Delta {\xi_{\rm t}}$ (km\,s$^{-1}$) \\
 & ($V-K$)  & ($V-K$) & (Spec) & (Spec) & (Parallax) & (Parallax) & (Spec) & (Spec) & & & & & \\
  \hline\noalign{\smallskip}
HD\,77912  & 4995 & 304 & 5021 & 59 & 1.99 & 0.11 & 2.06 & 0.10 &    0.05   & 0.13 &    0.05 & 2.05 & 0.16 \\
HD\,85503  & 4493 & 124 & 4635 & 70 & 2.62 & 0.07 & 2.65 & 0.13 &    0.38   & 0.17 &    0.36 & 1.50 & 0.15 \\
HD\,92482  & 4778 & 33  & 4900 & 46 & 2.34 & 0.03 & 2.36 & 0.08 & $-$0.24   & 0.15 & $-$0.22 & 1.50 & 0.10 \\
HD\,101079 & 4982 & 40  & 5100 & 67 & 2.70 & 0.02 & 2.90 & 0.12 &    0.25   & 0.15 &    0.25 & 1.40 & 0.15 \\
HD\,107950 & 5184 & 359 & 5247 & 63 & 2.62 & 0.13 & 2.55 & 0.10 &    0.23   & 0.13 &    0.23 & 1.60 & 0.13 \\
HD\,115927 & 4581 & 26  & 4881 & 40 & 2.29 & 0.05 & 2.68 & 0.08 & $-$0.21   & 0.15 & $-$0.19 & 1.25 & 0.18 \\
HD\,119650 & 4450 & 16  & 4510 & 38 & 1.83 & 0.03 & 2.05 & 0.12 & $-$0.22   & 0.15 & $-$0.22 & 1.52 & 0.15 \\
HD\,125079 & 5351 & 36  & 5334 & 50 & 3.15 & 0.02 & 3.20 & 0.15 & $-$0.16   & 0.14 & $-$0.16 & 1.20 & 0.15 \\
HD\,130255 & 4436 & 24  & 4509 & 72 & 1.71 & 0.06 & 1.90 & 0.15 & $-$0.86   & 0.14 & $-$0.88 & 1.30 & 0.16 \\
HD\,131873 & 4007 & 75  & 4150 & 80 & 1.40 & 0.06 & 1.40 & 0.15 & $-$0.08   & 0.14 & $-$0.09 & 1.65 & 0.10 \\
HD\,133208 & 4931 & 166 & 5174 & 40 & 2.52 & 0.06 & 2.54 & 0.11 &    0.23   & 0.13 &    0.23 & 1.70 & 0.20 \\
HD\,136138 & 5123 & 375 & 5100 & 33 & 2.76 & 0.14 & 2.92 & 0.05 &    0.07   & 0.15 &    0.10 & 1.45 & 0.17 \\
HD\,150430 & 4891 & 31  & 4850 & 40 & 2.60 & 0.03 & 2.57 & 0.05 & $-$0.26   & 0.13 & $-$0.21 & 1.35 & 0.14 \\
HD\,151101 & 4482 & 244 & 4504 & 34 & 1.78 & 0.12 & 1.81 & 0.06 &    0.13   & 0.14 &    0.16 & 1.70 & 0.05 \\
HD\,160507 & 4936 & 48  & 5044 & 47 & 2.85 & 0.03 & 2.91 & 0.07 &    0.09   & 0.14 &    0.13 & 1.40 & 0.14 \\
HD\,160538 & 4572 & 340 & 4812 & 89 & 2.75 & 0.16 & 3.00 & 0.15 & $-$0.02   & 0.13 & $-$0.05 & 1.70 & 0.20 \\
HD\,168532 & 4293 & 190 & 4307 & 39 & 1.30 & 0.10 & 1.90 & 0.08 &    0.33   & 0.16 &    0.34 & 1.65 & 0.08 \\
HD\,176524 & 4621 & 270 & 4700 & 24 & 2.19 & 0.12 & 2.20 & 0.07 &    0.13   & 0.15 &    0.15 & 1.55 & 0.12 \\
HD\,176670 & 4067 & 118 & 4300 & 52 & 1.05 & 0.08 & 1.35 & 0.18 &    0.09   & 0.14 &    0.06 & 1.80 & 0.05 \\
HD\,177304 & 5093 & 32  & 5200 & 41 & 2.34 & 0.02 & 2.60 & 0.10 &    0.23   & 0.14 &    0.24 & 1.65 & 0.15 \\
  \noalign{\smallskip}\hline
\end{tabular}
}
\end{center}
\end{table*}
\begin{table}
\begin{center}
\caption[]{Physical parameters of the sample stars (continued).}
\label{tab:parameter2}
\begin{tabular}{lcc}
  \hline\noalign{\smallskip}
Star name & $M$/M$_{\odot}$ & $\log$\,$L$/L$_{\odot}$ \\
  \hline\noalign{\smallskip}
HD\,77912  & 4.50 & 2.79 \\
HD\,85503  & 1.30 & 1.52 \\
HD\,92482  & 1.90 & 2.07 \\
HD\,101079 & 3.00 & 1.80 \\
HD\,107950 & 3.00 & 2.20 \\
HD\,115927 & 0.90 & 1.42 \\
HD\,119650 & 1.60 & 2.16 \\
HD\,125079 & 2.20 & 1.44 \\
HD\,130255 & 0.70 & 1.95 \\
HD\,131873 & 1.80 & 2.72 \\
HD\,133208 & 3.50 & 2.25 \\
HD\,136138 & 3.00 & 1.78 \\
HD\,150430 & 1.85 & 1.83 \\
HD\,151101 & 4.50 & 2.85 \\
HD\,160507 & 2.20 & 1.63 \\
HD\,160538 & 1.00 & 1.12 \\
HD\,168532 & 6.00 & 2.81 \\
HD\,176524 & 2.50 & 2.28 \\
HD\,176670 & 4.50 & 3.23 \\
HD\,177304 & 3.50 & 2.20 \\
  \noalign{\smallskip}\hline
\end{tabular}
\end{center}
\end{table}

The surface gravities were also derived using two methods. The first method
is based on the formulas as follows:
\begin{equation}
\label{eq:logg}
\log {\frac{g}{g_{\odot}}}=\log {\frac{M}{M_{\odot}}}
+4\log {\frac{T_{\rm eff}}{T_{{\rm eff}\odot}}}+0.4(M_{\rm bol}-M_{{\rm bol}\odot})
\end{equation}
\begin{equation}
\label{eq:mass}
M_{\rm bol}=V-A_{V}+5+5\log \pi +BC,
\end{equation}
here, $M$, $M_{\rm bol}$, $V$, $A_{V}$, $\pi$, and $BC$ represent the stellar mass, absolute bolometric
magnitude, apparent magnitude, interstellar extinction, parallax and bolometric correction, respectively.
The stellar mass $M$ was derived from the $Y^{2}$ (Yonsei-Yale) Stellar Models \citep{Yi2003}. The parallaxes of
HD\,131873 and the other sample stars were adopted from {\it Hipparcos} and {\it Gaia}\,DR3 \citep{Gaia2023},
respectively. The bolometric correction $BC$ was calculated using the relation provided by \citet{Alonso1999}.
For the Sun, $T_{{\rm eff}\odot}$, $\log$\,$g_{\odot}$ and $M_{{\rm bol}\odot}$ were adopted as 5770\,K,
4.44\,dex and 4.75\,mag, respectively. The second method for determining $\log$\,$g$ involves using the ionization
equilibrium. We obtained the final $\log$\,$g$ with this method by requiring that the difference of iron
abundances from \ion {Fe}{i} and \ion {Fe}{ii} lines is less than 0.05\,dex. The corresponding
uncertainties in surface gravities were estimated using the two formulas mentioned above and by adjusting
$\log$\,$g$ to force the abundance difference between \ion{Fe}{i} and \ion{Fe}{ii} lines to reach $\pm$0.05\,dex.

The micro-turbulent velocities were determined by adjusting them until the
\ion {Fe}{i} abundances exhibited no dependence on EW. The uncertainties were estimated by adjusting the
$\xi_{\rm t}$ to force the slope of the linear fit to be 0.001. Metallicities were initially either
adopted from previous papers or set to be 0.0\,dex. The final results were obtained by iteratively adjusting all
the four atmospheric parameters mentioned above until they reached a consistent solution. The uncertainties
in [Fe/H] were derived from the scatter of iron abundances measured from different \ion {Fe}{i} lines.

\begin{figure}
\begin{center}
\includegraphics[width=\columnwidth]{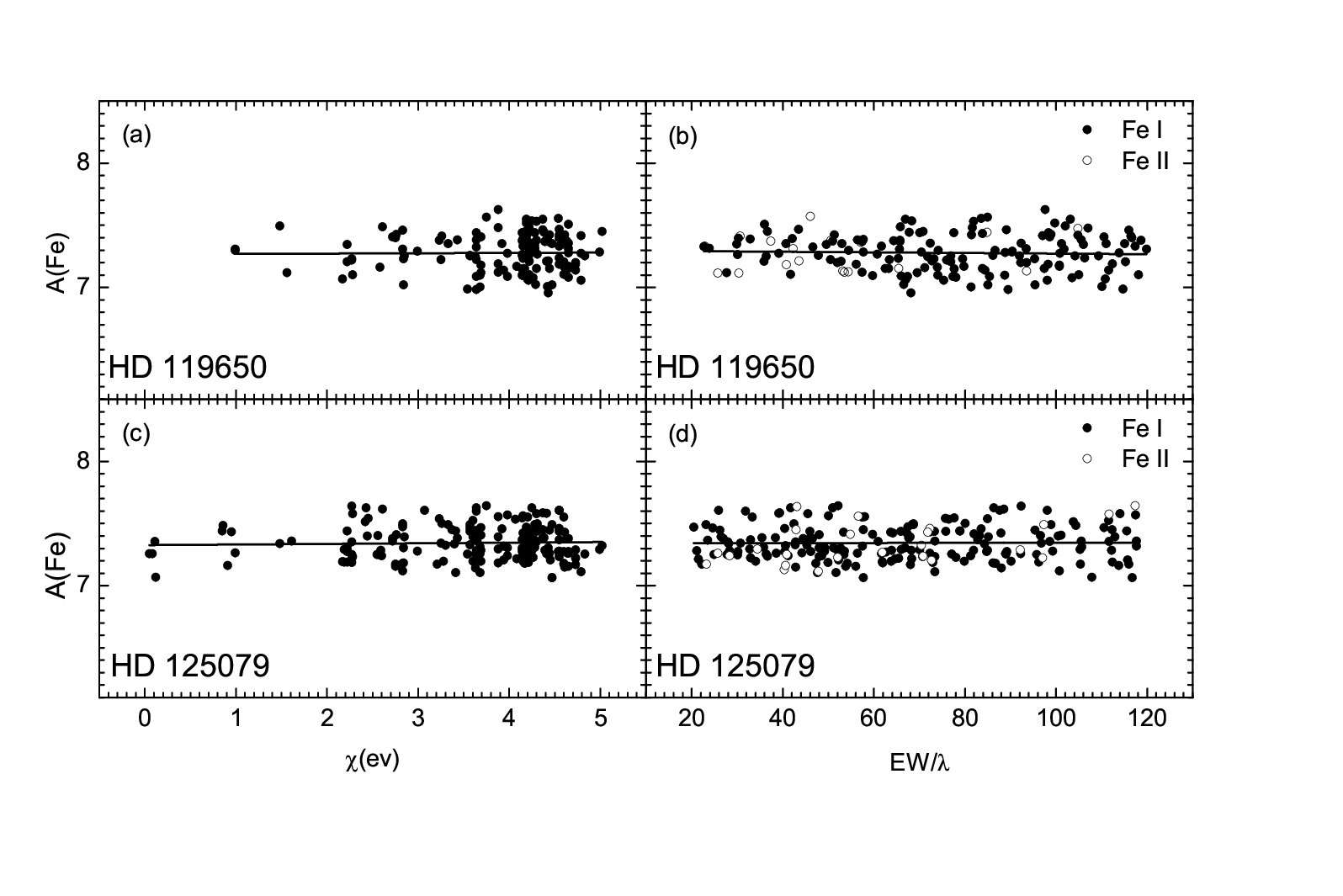}
\vspace{-1cm}
\caption{The determination of atmospheric parameters of the sample stars HD\,119650 (the upper two panels)
and HD\,125079 (the lower two panels) based on the spectra. The filled and open circles refer to the Fe
abundances computed from \ion {Fe}{i} and \ion {Fe}{ii} lines, respectively. The straight lines are the
linear-fitting results of \ion {Fe}{i} abundances. Panels (a) and (c) show the excitation equilibrium of Fe
abundances derived from \ion {Fe}{i} lines. Panels (b) and (d) show the ionization equilibrium of Fe abundances
from \ion {Fe}{i} and \ion {Fe}{ii} lines. Furthermore, the EW-independence of \ion {Fe}{i} abundances is also
shown in panels (b) and (d).}
\label{fig:parameter}
\end{center}
\end{figure}

Taking HD\,119650 and HD\,125079 as examples, Fig.\,\ref{fig:parameter} shows the excitation
equilibrium, ionization equilibrium and EW-independent of the derived iron abundances from \ion {Fe}{i}
and \ion {Fe}{ii} lines. The stellar atmospheric parameters derived from different methods for the 20 sample stars
are provided in Table\,\ref{tab:parameter}. Furthermore, the stellar masses derived from the $Y^2$ models
and the luminosities of our sample stars calculated based on masses are listed in Table\,\ref{tab:parameter2}.
Based on the analysis of stellar atmosphere parameters above, we estimated
the typical uncertainties of the stellar atmosphere parameters for our sample stars as $\Delta {T_{\rm eff}}$ = 100\,K,
$\Delta {\log g}$ = 0.1\,dex, $\Delta {\rm [Fe/H]}$ = 0.1\,dex and $\Delta {\xi_{\rm t}}$ = 0.2\,km\,s$^{-1}$. These
uncertainties are later used in the abundance uncertainty estimations (see Sect.\,4.1).
From the table, we can see that $T_{\rm eff}$ of HD\,115927
estimated by $V-K$ is 300\,K lower than that derived from the spectroscopic method. Based on the photometric colors,
\citet{Casagrande2011} computed the $T_{\rm eff}$ of this star as 4566\,K, which is similar to the value derived from
$V-K$ in this work. By fitting synthetic spectra, \citet{Forsberg2019} computed the $T_{\rm eff}$ of this star as
4768\,K with an uncertainty of 50\,K. Using the spectroscopic method, we derived the $T_{\rm eff}$ of this star as
4881\,K with an uncertainty of 40\,K, which is close to the value of \citet{Forsberg2019} when
considering the uncertainties. Additionally, the $\log$\,$g$ of HD\,168532 obtained through the parallax of
{\it Gaia} DR3 is 0.60\,dex smaller than that obtained by the spectroscopic method. The reasons for this discrepancy
may include: Firstly, the extinction of the V magnitude of this star is significant, reaching 0.56. Secondly, the relative
error in the parallax of this star is large, reaching 5.9\%.

For checking the reliability of our derived stellar atmospheric parameters, we present a comparison of our obtained stellar
atmospheric parameters with those from the literature, whenever available, in Table\,\ref{tab:compare-parameter}.
The mean differences of the four stellar atmospheric parameters for the common stars between this work and the literature are
$\Delta T_{\rm eff} \lesssim$ 123\,K, $\Delta\log$\,$g$ $\lesssim$ 0.24\,dex, $\Delta$[Fe/H] $\lesssim 0.20$\,dex
and $\Delta \xi_{\rm t} \lesssim 0.28$\,km\,s$^{-1}$. For HD\,119650, a significant discrepancy between our
values and those of \citet{Mahanta2016} can be seen. Both our study and \citet{Mahanta2016} employed the
spectroscopic method to determine these parameters, and the difference in [Fe/H] may be attributed to the use of
different Fe lines and atomic data (e.g., $\log$\,$gf$). \citet{Mahanta2016} measured the Fe abundance
of this star using 101 \ion {Fe}{i} and 7 \ion {Fe}{ii} lines, while we utilized 142 \ion {Fe}{i} and 18 \ion {Fe}{ii} lines.
The $T_{\rm eff}$ of this star in our work is 315\,K lower than that of \citet{Mahanta2016},
and the $\log$\,$g$ is 0.80\,dex lower. We notice that the $T_{\rm eff}$ and $\log$\,$g$ values of this star
determined through color indices $V-K$ and parallax are 4450\,K and 1.83\,dex, respectively, which are very close
to our results. There is also a significant disparity in the $\log$\,$g$ of HD\,176670 between this study and
\citet{McWilliam1990}, which may be attributed to our use of the reliable Gaia parallax data,
which were not available in 1990 when neither {\it Hipparcos} nor {\it Gaia} data existed.

\begin{table}
\begin{center}
\scriptsize
\caption{Comparisons of atmospheric parameters with literature values.}
\label{tab:compare-parameter}
\setlength\tabcolsep{2pt}
\begin{threeparttable}
\begin{tabular}{lrcrcc}
  \hline\noalign{\smallskip}
Star name & $T_{\rm eff}$ & $\log$\,$g$ & [Fe/H] & $\xi_{\rm t}$ & References \\
  \hline\noalign{\smallskip}
HD\,77912   & 5021 & 2.06 &  0.05   & 2.05 & 1 \\
            & 4899 & 1.75 & $-$0.14 & 2.13 & 2 \\
            & 5001 & 2.03 &  0.12  & 2.16  & 3 \\
HD\,85503   & 4635 & 2.65 &  0.38  & 1.50  & 1 \\
            & 4480 & 2.61 &  0.17  & 2.20  & 4 \\
HD\,101079  & 5100 & 2.90 &  0.25   & 1.40 & 1 \\
            & 5000 & 2.70 &  0.10   & 1.30 & 5 \\
HD\,107950  & 5247 & 2.55 &  0.23   & 1.60 & 1 \\
            & 5171 & 2.60 &  0.01   & 1.63 & 2 \\
HD\,115927  & 4881 & 2.68 &  $-$0.21 & 1.25 & 1 \\
            & 4768 & 2.64 &  $-$0.50 & 1.29 & 6 \\
HD\,119650  & 4510 & 2.05 & $-$0.22 & 1.52 & 1 \\
            & 4825 & 2.85 &  0.04   & 1.62 & 7 \\
            & 4500 & 1.60 & $-$0.10 & 1.40 & 8 \\
HD\,125079  & 5334 & 3.20 & $-$0.16 & 1.20 & 1 \\
            & 5520 & 3.30 & $-$0.18 & 1.25 & 9 \\
HD\,130255  & 4509 & 1.90 & $-$0.88 & 1.30 & 1 \\
            & 4400 & 1.50 & $-$1.11 & 1.30 & 8 \\
HD\,131873  & 4150 & 1.40 & $-$0.08 & 1.65 & 1 \\
            & 4030 & 1.83 & $-$0.29 & 2.40 & 4 \\
HD 133208   & 5174 & 2.54 &  0.23   & 1.70 & 1 \\
            & 5001 & 2.35 & $-$0.07 & 1.61 & 2 \\
HD\,136138  & 5100 & 2.92 &   0.07  & 1.45 & 1 \\
            & 4995 & 2.80 & $-$0.19 & 1.50 & 10 \\
HD\,160507  & 5044 & 2.91 & 0.09    & 1.40 & 1 \\
            & 5029 & 2.96 & $-$0.04 & 1.57 & 11 \\
HD\,160538  & 4812 & 3.00 & $-$0.02 & 1.70 & 1 \\
            & 4851 & 3.00 & $-$0.14 & 1.76 & 12 \\
HD\,168532  & 4307 & 1.90 & 0.33    & 1.65 & 1 \\
            & 4130 & 2.05 & 0.00    & 2.40 & 4 \\
HD\,176524  & 4700 & 2.20 & 0.13    & 1.55 & 1 \\
            & 4520 & 2.55 & $-$0.12 & 2.10 & 4 \\
HD\,176670  & 4300 & 1.35 & 0.09    & 1.80 & 1 \\
            & 4220 & 2.21 & $-$0.03 & 2.80 & 4 \\
\noalign{\smallskip}\hline\\
\end{tabular}
\begin{tablenotes}
\item References: 1. This work; 2. \citet{Takeda2008}; 3. \citet{Luck2014}; 4. \citet{McWilliam1990}; 5. \citet{Pereira2011};
6. \citet{Forsberg2019}; 7.\citet{Mahanta2016}; 8. \citet{deCastro2016}; 9. \citet{Karinkuzhi2014}; 10. \citet{Mishenina2006};
11. \citet{Wang2011}; 12. \citet{Merle2016}.
\end{tablenotes}
\end{threeparttable}
\end{center}
\end{table}


\section{Analysis of chemical abundances}
\subsection{Abundances and their uncertainties}

Based on the EWs and stellar atmospheric parameters derived above, we determined the abundances of light odd-Z elements
(Na and Al), $\alpha$-elements (Mg, Si, Ca and Ti), Fe-peak elements (Cr, Fe and Ni),
and n-capture elements (Sr, Y, Zr, Ce, Nd and Sm). Our methodology involved using the model atmospheres, which were interpolated from a
grid of plane-parallel, local thermodynamic equilibrium (LTE) models provided by \citet{Kurucz1993}. Additionally, we employed
the ABONTEST8 program, developed by Dr. P. Magain, to calculate the theoretical line EWs. The final determination of abundances
was made by ensuring consistency between the theoretical EWs and observed values. The entire set of abundance-analysis procedures
used in this work has been employed in other studies \citep[e.g.,][]{Liang2003,Liu2009,Yang2016,Kong2018a}. For the elements
Sc, Mn, Ba, La and Eu, we determined the abundances using the spectral synthesis method. We utilized the IDL code
Spectrum Investigation Utility (SIU) developed by \citet{Reetz1999} to fit the observed spectra. The final abundances of the
20 sample stars are listed in Tables\,\ref{tab:abun1} and \ref{tab:abun2}. The chemical abundances of the Sun were adopted
from \citet{Asplund2009}.

\begin{figure}
\centering
\includegraphics[width=\columnwidth]{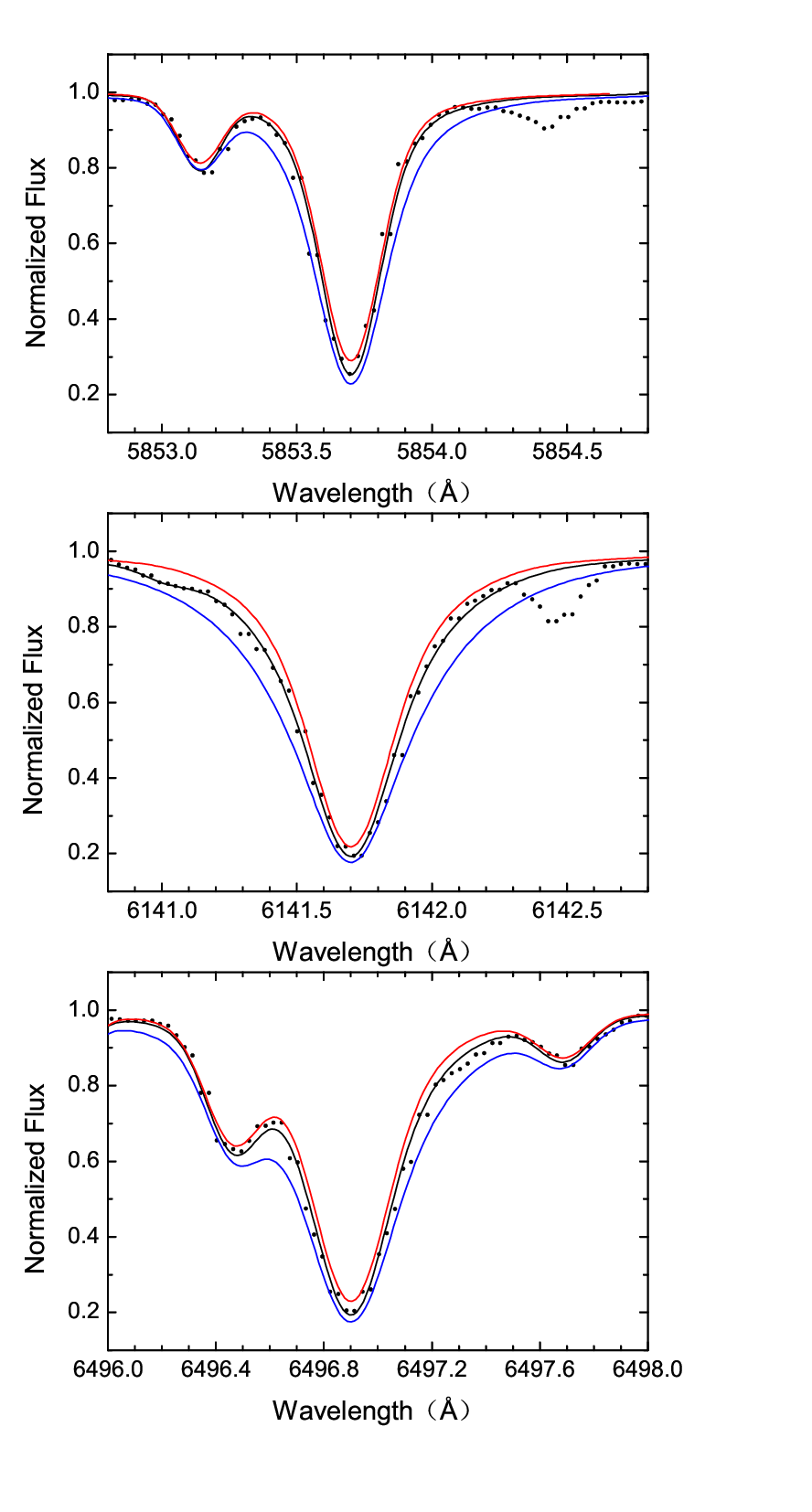}
\vspace{-1cm}
\caption{Examples of spectral fittings of Ba lines at 5853.67, 6141.71 and 6496.91\,\AA\, for HD\,92482. The observed and
synthetic spectra are shown with filled circles and black lines respectively. The blue and red lines are the synthetic
spectra for $\Delta\log\epsilon(\mathrm{Ba}) = \pm$ 0.2\,dex around the best fits.}
\label{fig:syn}
\end{figure}

The atomic data for most of the observed lines are adopted from the NIST\footnote{http://physics.nist.gov/asd} database
\citep{Kramida2023}, as well as from \citet{Bensby2014} and \citet{desmedt2015}. Additionally, we supplemented our data for the
20 elements with information from other sources. Detailed information is provided in Table \ref{tab:atomic-data}. Furthermore,
we considered the hyperfine structure (HFS) that splits the atomic terms into hyperfine components and desaturates the spectral
lines, especially for the strong ones. Taking the mild Ba-star HD\,119650 and the strong Ba-star HD\,125079 as examples, the EWs
of the Ba lines at 5853.67, 6141.71 and 6496.91\,\AA~in HD\,119650 were measured as 154.6, 229.1 and 233.0\,m\AA,
respectively. For HD\,125079, these values are 149.4, 255.6 and 235.5\,m\AA. We accounted for the HFS of Sc \citep{Lawler2019},
Mn \citep{Den2011}, Ba \citep{McWilliam1998}, La \citep{Ivans2006} and Eu \citep{Lawler2001} in this work. Additionally,
considering that the non-local thermodynamic equilibrium (NLTE) effects are significant for Ba abundances
\citep{Mashonkina1999,Reddy2017,Liu2021}, the NLTE effects have been considered in determining the Ba abundances.
Fig. \ref{fig:syn} shows the fitting for the three \ion {Ba}{ii} lines of HD\,92482.

\begin{table}
\begin{center}
\caption[]{Absorption lines used for abundance determination.}
\label{tab:atomic-data}
\begin{threeparttable}
\begin{tabular}{lrrrc}
  \hline\noalign{\smallskip}
Species & $\lambda($\AA$)$ & $\chi(eV)$ & $\log$\,$gf$ & Ref. \\
  \hline\noalign{\smallskip}
\ion{Na}{i} & 5682.63 & 2.10 & $-$0.71 & 1 \\
            & 5688.21 & 2.10 & $-$0.45 & 1 \\
            & 6154.23 & 2.10 & $-$1.55 & 1 \\
            & 6160.75 & 2.10 & $-$1.25 & 1 \\
\ion{Mg}{i} & 4571.10 & 0.00 & $-$5.47 & 2 \\
            & 4703.00 & 4.34 & $-$0.37 & 2 \\
            & 5528.42 & 4.34 & $-$0.47 & 2 \\
            & 5711.10 & 4.34 & $-$1.75 & 2 \\
            & 6799.00 & 5.75 & $-$1.56 & 2 \\
\ion{Al}{i} & 5557.07 & 3.14 & $-$2.23 & 3 \\
            & 6696.03 & 3.14 & $-$1.35 & 4 \\
            & 6698.67 & 3.14 & $-$1.65 & 4 \\
\noalign{\smallskip}\hline
\end{tabular}
\begin{tablenotes}
\item References: 1. NIST; 2. \citet{Zhao2016}; 3. \citet{Bensby2014}; 4. \citet{desmedt2015}; 5. \citet{Lawler2019};
6. \citet{Roederer2018}; 7. \citet{Den2011}; 8. \citet{OBrian1991}; 9. \citet{Fuhr1988}; 10. \citet{Lambert1996};
11. \citet{Bard1994}; 12. \citet{Chen2000}; 13. \citet{Nissen1997}; 14. \citet{Hannaford1982}; 15. \citet{Biemont1981};
16. \citet{McWilliam1998}; 17. \citet{Mashonkina2000}; 18. \citet{Lawler2001}.
\item (The complete version of this table is available in machine-readable form in the online edition of the journal.)
\end{tablenotes}
\end{threeparttable}
\end{center}
\end{table}

In our study, we have determined detailed elemental abundances for four stars (i.e., HD\,92482, HD\,150430, HD\,151101
and HD\,177304) for the first time. As for the remaining 16 sample stars, four of them (HD\,101079, HD\,115927, HD\,130255 and
HD\,176670) have had their chemical abundances analyzed by \citet{Pereira2011}, \citet{Forsberg2019}, \citet{deCastro2016} and
\citet{McWilliam1990}, respectively, but none provided the Ba abundances. \citet{Fernandez1990} analyzed Ba and Fe abundances
for two stars (HD\,77912 and HD\,133208) using intermediate dispersion spectra. Hence, the Ba abundances for these six stars
have been derived from high-resolution spectra for the first time. While the Ba abundances of the remaining 10 sample stars in
our study have been explored before \cite[e.g.,][]{Zacs1997,Luck2015,Kong2018b,Casamiquela2020}, most researchers used EW
measurements to obtain the abundance results. In this work, we employed the spectral synthesis method to calculate the Ba abundance
taking into account both the HFS and NLTE effects. Using HD\,119650 and HD\,125079 as examples, the standard deviations of
Ba abundances from line to line are approximately 0.05 and 0.06\,dex, respectively. The NLTE corrections
$\Delta_{\mathrm{NLTE}}$ (= $\log\epsilon(\mathrm{Ba})_{\mathrm{NLTE}} - \log\epsilon(\mathrm{Ba})_{\mathrm{LTE}}$)
for different lines vary. The largest absolute values of the corrections for the two stars both correspond to the line at
6496.91\,\AA\,, reaching $\Delta_{\mathrm{NLTE}} = -0.16$ and $-$0.33\,dex, respectively. For comparison, the LTE and NLTE
Ba abundances of the sample stars are presented in Tables\,\ref{tab:abun1} and \ref{tab:abun2}. Our results demonstrate that NLTE analyses are critical for enhancing
the accuracy of stellar abundance measurements.

\begin{table*}
\centering \scriptsize \caption{Element abundances of the sample stars.}
\label{tab:abun1} \setlength\tabcolsep{2pt}
\begin{threeparttable}
\begin{tabular}{lrrrrrrrrrrrrrrrrrrrrrrrrr}
\hline \noalign{\smallskip} &&
\multicolumn{2}{c}{HD\,77912} && \multicolumn{2}{c}{HD\,85503} && \multicolumn{2}{c}{HD\,92482}
&& \multicolumn{2}{c}{HD\,101079} && \multicolumn{2}{c}{HD\,107950} && \multicolumn{2}{c}{HD\,115927} && \multicolumn{2}{c}{HD\,119650}\\
\cline{3-4} \cline{6-7} \cline{9-10} \cline{12-13} \cline{15-16} \cline{18-19} \cline{21-22} \\
Species && $\log\epsilon$(X) & [X/Fe] && $\log\epsilon$(X) & [X/Fe] &&
$\log\epsilon$(X) & [X/Fe] && $\log\epsilon$(X) & [X/Fe] &&
$\log\epsilon$(X) & [X/Fe] && $\log\epsilon$(X) & [X/Fe] &&
$\log\epsilon$(X) & [X/Fe]  \\
\noalign{\smallskip} \hline \noalign{\smallskip}
\ion{Fe}{i}  && 7.55 &      -- && 7.88 &      -- && 7.26 &      -- && 7.75 &      -- && 7.73 &      -- && 7.29 &      -- && 7.28 &      -- \\
\ion{Fe}{ii} && 7.55 &      -- && 7.86 &      -- && 7.28 &      -- && 7.75 &      -- && 7.73 &      -- && 7.31 &      -- && 7.28 &      -- \\
\ion{Na}{i}  && 6.78 &    0.49 &&   -- &     --  && 6.27 &    0.27 && 6.75 &    0.26 && 6.78 &    0.31 && 6.14 &    0.11 && 6.25 &    0.23 \\
\ion{Mg}{i}  && 7.59 & $-$0.06 && 7.92 & $-$0.06 && 7.43 &    0.07 && 7.83 & $-$0.02 && 7.75 & $-$0.08 && 7.40 &    0.01 && 7.30 & $-$0.08 \\
\ion{Al}{i}  && 6.57 &    0.07 && 7.06 &    0.23 && 6.19 & $-$0.02 && 6.66 & $-$0.04 && 6.60 & $-$0.08 && 6.14 & $-$0.10 && 6.15 & $-$0.08 \\
\ion{Si}{i}  && 7.64 &    0.08 && 8.17 &    0.28 && 7.56 &    0.29 && 7.83 &    0.07 && 7.78 &    0.04 && 7.44 &    0.14 && 7.51 &    0.22 \\
\ion{Ca}{i}  && 6.41 &    0.02 && 6.73 &    0.01 && 6.13 &    0.03 && 6.55 & $-$0.04 && 6.61 &    0.04 && 6.13 &    0.00 && 6.04 & $-$0.08 \\
\ion{Sc}{ii} && 3.31 &    0.11 && 3.69 &    0.16 && 3.06 &    0.15 && 3.56 &    0.16 && 3.45 &    0.07 && 3.07 &    0.13 && 3.11 &    0.18 \\
\ion{Ti}{i}  && 4.96 & $-$0.04 && 5.43 &    0.10 && 4.69 & $-$0.02 && 5.13 & $-$0.07 && 5.11 & $-$0.07 && 4.78 &    0.04 && 4.60 & $-$0.13 \\
\ion{Ti}{ii} && 5.35 &    0.35 && 5.47 &    0.14 && 4.95 &    0.24 && 5.36 &    0.16 && 5.44 &    0.26 && 5.09 &    0.35 && 4.82 &    0.09 \\
\ion{Cr}{i}  && 5.84 &    0.15 && 6.08 &    0.06 && 5.43 &    0.03 && 5.86 & $-$0.03 && 5.97 &    0.10 && 5.38 & $-$0.05 && 5.28 & $-$0.14 \\
\ion{Cr}{ii} && 6.16 &    0.47 && 6.42 &    0.40 && 5.78 &    0.38 && 6.09 &    0.20 && 6.22 &    0.35 && 5.71 &    0.28 && 5.59 &    0.17 \\
\ion{Mn}{i}  && 5.43 & $-$0.06 && 5.80 & $-$0.02 && 5.30 &    0.11 && 5.73 &    0.05 && 5.62 & $-$0.05 && 5.25 &    0.03 && 5.23 &    0.02 \\
\ion{Ni}{i}  && 6.26 & $-$0.01 && 6.81 &    0.21 && 6.00 &    0.02 && 6.50 &    0.03 && 6.43 & $-$0.02 && 6.01 &    0.00 && 5.99 & $-$0.01 \\
\ion{Sr}{i}  &&   -- &     --  &&   -- &     --  && 3.46 &    0.83 && 3.26 &    0.14 && 2.94 & $-$0.16 && 3.08 &    0.42 &&   -- &     --  \\
\ion{Y}{i}   && 2.53 &    0.27 && 2.55 & $-$0.04 && 2.75 &    0.78 && 2.82 &    0.36 &&   -- &     --  && 2.48 &    0.48 && 2.03 &    0.04 \\
\ion{Y}{ii}  && 2.72 &    0.46 && 3.02 &    0.43 && 3.23 &    1.26 && 3.17 &    0.71 && 2.73 &    0.29 && 2.84 &    0.84 && 2.65 &    0.66 \\
\ion{Zr}{i}  && 2.82 &    0.19 && 2.91 & $-$0.05 && 3.20 &    0.86 && 3.15 &    0.32 &&   -- &     --  && 2.74 &    0.37 && 2.40 &    0.04 \\
\ion{Zr}{ii} && 3.15 &    0.52 && 3.54 &    0.58 && 3.61 &    1.27 && 3.42 &    0.59 && 3.04 &    0.23 && 3.14 &    0.77 && 2.98 &    0.62 \\
\ion{Ba}{ii}               && 2.86 &    0.63 && 2.87 &    0.31 && 3.46 &    1.52 && 3.05 &    0.62 && 2.72 &    0.31 && 3.11 &    1.14 && 2.61 &    0.65 \\
\ion{Ba}{ii}\tnote{$\ast$} && 2.79 &    0.56 && 2.81 &    0.25 && 3.38 &    1.44 && 2.91 &    0.48 && 2.67 &    0.26 && 3.04 &    1.07 && 2.54 &    0.58 \\
\ion{La}{ii} && 1.40 &    0.25 && 1.83 &    0.35 && 2.30 &    1.44 && 1.79 &    0.44 && 1.45 &    0.12 && 2.04 &    1.15 && 1.48 &    0.60 \\
\ion{Ce}{ii} && 1.92 &    0.29 && 2.37 &    0.41 && 2.83 &    1.49 && 2.36 &    0.53 && 2.18 &    0.37 && 2.70 &    1.33 && 1.69 &    0.33 \\
\ion{Nd}{ii} && 1.96 &    0.49 && 2.26 &    0.46 && 2.60 &    1.42 && 2.37 &    0.70 && 2.17 &    0.52 && 2.57 &    1.36 && 1.84 &    0.64 \\
\ion{Sm}{ii} && 1.52 &    0.51 && 1.88 &    0.54 && 2.00 &    1.28 && 1.57 &    0.36 && 1.43 &    0.24 && 2.04 &    1.29 && 1.09 &    0.35 \\
\ion{Eu}{ii} && 0.72 &    0.15 && 1.00 &    0.10 && 0.85 &    0.57 && 0.80 &    0.03 && 0.77 &    0.02 && 0.63 &    0.32 && 0.60 &    0.30 \\
\noalign{\smallskip} \hline
\end{tabular}
\centering \scriptsize
\setlength\tabcolsep{2pt}
\begin{tabular}{lrrrrrrrrrrrrrrrrrrrrrrrrr}
\hline \noalign{\smallskip} &&
\multicolumn{2}{c}{HD\,125079} && \multicolumn{2}{c}{HD\,130255} && \multicolumn{2}{c}{HD\,131873} &&
\multicolumn{2}{c}{HD\,133208} && \multicolumn{2}{c}{HD\,136138} && \multicolumn{2}{c}{HD\,150430} && \multicolumn{2}{c}{HD\,151101} \\
\cline{3-4} \cline{6-7} \cline{9-10} \cline{12-13} \cline{15-16} \cline{18-19} \cline{21-22} \\
Species && $\log\epsilon$(X) & [X/Fe] && $\log\epsilon$(X) & [X/Fe] &&
$\log\epsilon$(X) & [X/Fe] && $\log\epsilon$(X) & [X/Fe] &&
$\log\epsilon$(X) & [X/Fe] && $\log\epsilon$(X) & [X/Fe] &&
$\log\epsilon$(X) & [X/Fe]  \\
\noalign{\smallskip} \hline \noalign{\smallskip}
\ion{Fe}{i}  && 7.34 &      -- && 6.64    &      -- && 7.42 &      -- && 7.73 &      -- && 7.57 &      -- && 7.24 &      -- && 7.63 &      -- \\
\ion{Fe}{ii} && 7.34 &      -- && 6.62    &      -- && 7.41 &      -- && 7.73 &      -- && 7.60 &      -- && 7.29 &      -- && 7.66 &      -- \\
\ion{Na}{i}  && 6.20 &    0.12 && 5.41    &    0.03 && 6.46 &    0.30 && 6.77 &    0.30 && 6.47 &    0.16 && 6.22 &    0.24 && 6.66 &    0.29 \\
\ion{Mg}{i}  && 7.44 &    0.00 && 6.92    &    0.18 && 7.52 &    0.00 &&   -- &      -- && 7.62 & $-$0.05 && 7.32 & $-$0.02 && 7.77 &    0.04 \\
\ion{Al}{i}  && 6.07 & $-$0.22 && 5.83    &    0.24 && 6.49 &    0.12 && 6.68 &    0.00 && 6.60 &    0.08 && 6.18 & $-$0.01 && 6.54 & $-$0.04 \\
\ion{Si}{i}  && 7.42 &    0.07 && 7.18    &    0.53 && 7.52 &    0.09 && 7.84 &    0.10 && 7.68 &    0.10 && 7.43 &    0.18 && 7.75 &    0.11 \\
\ion{Ca}{i}  && 6.23 &    0.05 && 5.70    &    0.22 && 6.22 & $-$0.04 && 6.61 &    0.04 && 6.42 &    0.01 && 5.97 & $-$0.11 && 6.37 & $-$0.10 \\
\ion{Sc}{ii} && 3.18 &    0.19 && 2.50    &    0.21 && 3.13 &    0.06 && 3.51 &    0.13 && 3.32 &    0.10 && 3.17 &    0.28 && 3.43 &    0.15 \\
\ion{Ti}{i}  && 4.76 & $-$0.03 && 4.31    &    0.22 && 5.06 &    0.19 && 5.14 & $-$0.04 && 5.01 & $-$0.01 && 4.55 & $-$0.14 && 4.98 & $-$0.10 \\
\ion{Ti}{ii} && 4.91 &    0.12 && 4.49    &    0.40 && 5.11 &    0.24 && 5.35 &    0.17 && 5.24 &    0.22 && 4.97 &    0.28 && 5.42 &    0.34 \\
\ion{Cr}{i}  && 5.51 &    0.03 && 4.78    &    0.00 && 5.50 & $-$0.06 && 5.92 &    0.05 && 5.73 &    0.02 && 5.27 & $-$0.11 && 5.84 &    0.07 \\
\ion{Cr}{ii} && 5.69 &    0.21 && 5.14    &    0.36 && 5.67 &    0.11 && 6.33 &    0.46 && 5.89 &    0.18 && 5.83 &    0.45 && 6.30 &    0.53 \\
\ion{Mn}{i}  && 5.24 & $-$0.03 && 4.28    & $-$0.29 && 5.53 &    0.18 && 5.73 &    0.07 && 5.49 & $-$0.01 && 5.09 & $-$0.08 && 5.57 &    0.01 \\
\ion{Ni}{i}  && 6.05 & $-$0.01 && 5.42    &    0.06 && 6.22 &    0.08 && 6.45 &    0.00 && 6.28 & $-$0.01 && 5.95 & $-$0.01 && 6.42 &    0.07 \\
\ion{Sr}{i}  && 3.54 &    0.83 &&   --    &     --  &&   -- &     --  &&   -- &     --  && 3.43 &    0.49 &&   -- &     --  &&   -- &      -- \\
\ion{Y}{i}   && 2.94 &    0.89 &&   --    &     --  && 2.12 & $-$0.01 &&   -- &     --  && 2.90 &    0.62 && 2.02 &    0.07 &&   -- &      -- \\
\ion{Y}{ii}  && 3.02 &    0.97 && 1.79    &    0.44 && 2.35 &    0.22 && 2.66 &    0.22 && 3.04 &    0.76 && 2.37 &    0.42 && 2.42 &    0.08 \\
\ion{Zr}{i}  && 3.32 &    0.90 && 1.91    &    0.19 && 2.46 & $-$0.04 && 3.28 &    0.47 && 3.18 &    0.53 && 2.40 &    0.08 && 2.38 & $-$0.33 \\
\ion{Zr}{ii} && 3.51 &    1.09 && 2.23    &    0.51 &&   -- &     --  && 3.26 &    0.45 && 3.53 &    0.88 && 2.82 &    0.50 &&   -- &     --  \\
\ion{Ba}{ii}               && 3.33 &    1.31 && 1.96    &    0.64 && 2.52 &    0.42 && 2.83 &    0.42 && 2.84 &    0.59 && 2.39 &    0.47 && 2.78 &    0.47 \\
\ion{Ba}{ii}\tnote{$\ast$} && 3.17 &    1.15 && 1.90    &    0.58 && 2.47 &    0.37 && 2.74 &    0.33 && 2.78 &    0.53 && 2.33 &    0.41 && 2.67 &    0.36 \\
\ion{La}{ii} && 1.97 &    1.03 && 0.63    &    0.39 && 1.32 &    0.30 && 1.56 &    0.23 && 1.66 &    0.49 && 1.24 &    0.40 && 1.32 &    0.09 \\
\ion{Ce}{ii} && 2.48 &    1.06 && 1.34    &    0.62 && 1.77 &    0.27 && 2.23 &    0.42 && 2.02 &    0.37 && 1.79 &    0.47 && 2.06 &    0.35 \\
\ion{Nd}{ii} && 2.28 &    1.02 && 1.29    &    0.73 && 1.99 &    0.65 && 2.23 &    0.58 && 2.20 &    0.71 && 1.65 &    0.49 && 2.31 &    0.76 \\
\ion{Sm}{ii} && 1.41 &    0.61 && 0.71    &    0.61 &&   -- &     --  && 1.55 &    0.36 && 1.51 &    0.48 && 1.09 &    0.39 && 1.66 &    0.57 \\
\ion{Eu}{ii} && 0.70 &    0.34 && $-$0.10 &    0.24 && 0.54 &    0.10 && 0.83 &    0.08 && 0.76 &    0.17 && 0.58 &    0.32 && 0.67 &    0.02 \\
\noalign{\smallskip} \hline
\end{tabular}
\begin{tablenotes}
\item[$\ast$] The NLTE-corrected Ba abundances are listed here.
\end{tablenotes}
\end{threeparttable}
\end{table*}
\begin{table*}
\centering \scriptsize
\caption{Element abundances of the sample stars (continued).}
\label{tab:abun2}
\setlength\tabcolsep{2pt}
\begin{threeparttable}
\begin{tabular}{lrrrrrrrrrrrrrrrrrrrrrrr}
\hline \noalign{\smallskip} &&
\multicolumn{2}{c}{HD\,160507} && \multicolumn{2}{c}{HD\,160538} && \multicolumn{2}{c}{HD\,168532}
&& \multicolumn{2}{c}{HD\,176524} && \multicolumn{2}{c}{HD\,176670} && \multicolumn{2}{c}{HD\,177304} \\
\cline{3-4} \cline{6-7} \cline{9-10} \cline{12-13} \cline{15-16} \cline{18-19}\\
Species && $\log\epsilon$(X) & [X/Fe] && $\log\epsilon$(X) & [X/Fe] &&
$\log\epsilon$(X) & [X/Fe] && $\log\epsilon$(X) & [X/Fe] &&
$\log\epsilon$(X) & [X/Fe] && $\log\epsilon$(X) & [X/Fe] \\
\noalign{\smallskip} \hline \noalign{\smallskip}
\ion{Fe}{i}  && 7.59 &      -- && 7.48 &      -- && 7.83 &      -- && 7.63 &      -- && 7.59 &      -- && 7.73 &      --  \\
\ion{Fe}{ii} && 7.63 &      -- && 7.45 &      -- && 7.84 &      -- && 7.65 &      -- && 7.56 &      -- && 7.74 &      --  \\
\ion{Na}{i}  && 6.53 &    0.20 && 6.55 &    0.33 && 6.84 &    0.27 && 6.58 &    0.21 && 6.85 &    0.52 && 6.77 &    0.30  \\
\ion{Mg}{i}  && 7.62 & $-$0.07 && 7.44 & $-$0.14 && 7.89 & $-$0.04 && 7.57 & $-$0.16 && 7.84 &    0.15 && 7.68 & $-$0.15  \\
\ion{Al}{i}  && 6.51 & $-$0.03 && 6.62 &    0.19 && 6.78 &    0.00 && 6.54 & $-$0.04 && 6.61 &    0.07 && 6.52 & $-$0.16  \\
\ion{Si}{i}  && 7.77 &    0.17 && 7.48 & $-$0.01 && 7.90 &    0.06 && 7.71 &    0.07 && 7.79 &    0.19 && 7.73 & $-$0.01  \\
\ion{Ca}{i}  && 6.42 & $-$0.01 && 6.49 &    0.17 && 6.61 & $-$0.06 && 6.49 &    0.02 && 6.41 & $-$0.02 && 6.52 & $-$0.05  \\
\ion{Sc}{ii} && 3.43 &    0.19 && 3.28 &    0.15 && 3.64 &    0.16 && 3.26 & $-$0.02 && 3.36 &    0.12 && 3.49 &    0.11  \\
\ion{Ti}{i}  && 4.98 & $-$0.06 && 5.01 &    0.08 && 5.43 &    0.15 && 5.11 &    0.03 && 5.20 &    0.16 && 5.14 & $-$0.04  \\
\ion{Ti}{ii} && 5.24 &    0.20 && 5.25 &    0.32 && 5.49 &    0.21 && 5.17 &    0.09 && 5.29 &    0.25 && 5.41 &    0.23  \\
\ion{Cr}{i}  && 5.68 & $-$0.05 && 5.72 &    0.10 && 5.98 &    0.01 && 5.77 &    0.00 && 5.75 &    0.02 && 5.85 & $-$0.02  \\
\ion{Cr}{ii} && 6.10 &    0.37 && 6.11 &    0.49 &&   -- &     --  && 5.90 &    0.13 && 6.18 &    0.45 && 6.02 &    0.15  \\
\ion{Mn}{i}  && 5.48 & $-$0.04 && 5.45 &    0.04 && 5.92 &    0.16 && 5.57 &    0.01 && 5.66 &    0.14 && 5.64 & $-$0.03  \\
\ion{Ni}{i}  && 6.35 &    0.04 && 6.20 &    0.00 && 6.77 &    0.22 && 6.35 &    0.00 && 6.47 &    0.16 && 6.36 & $-$0.09  \\
\ion{Sr}{i}  && 3.18 &    0.22 &&   -- &     --  &&  --  &     --  &&   -- &     --  &&   -- &     --  && 3.13 &    0.03  \\
\ion{Y}{i}   && 2.77 &    0.47 && 2.44 &    0.25 && 2.70 &    0.16 &&   -- &     --  && 2.33 &    0.03 && 2.64 &    0.20  \\
\ion{Y}{ii}  && 3.18 &    0.88 &&   -- &     --  && 3.07 &    0.53 && 2.42 &    0.08 && 2.61 &    0.31 && 2.60 &    0.16  \\
\ion{Zr}{i}  && 3.07 &    0.40 && 2.72 &    0.16 && 3.07 &    0.16 && 2.75 &    0.04 && 2.67 &    0.00 && 3.07 &    0.26  \\
\ion{Zr}{ii} && 3.41 &    0.74 &&   -- &     --  &&  --  &     --  && 2.76 &    0.05 && 3.25 &    0.58 && 3.32 &    0.51  \\
\ion{Ba}{ii}               && 3.12 &    0.85 && 2.55 &    0.39 && 3.05 &    0.54 && 2.70 &    0.39 && 2.77 &    0.50 && 2.90 &    0.49 \\
\ion{Ba}{ii}\tnote{$\ast$} && 3.02 &    0.75 && 2.41 &    0.25 && 2.99 &    0.48 && 2.50 &    0.19 && 2.67 &    0.40 && 2.81 &    0.40  \\
\ion{La}{ii} && 1.95 &    0.76 && 1.22 &    0.14 && 1.80 &    0.37 && 1.44 &    0.21 && 1.38 &    0.19 && 1.54 &    0.21  \\
\ion{Ce}{ii} && 2.37 &    0.70 && 1.89 &    0.33 && 2.29 &    0.38 && 2.04 &    0.33 && 2.22 &    0.55 && 2.20 &    0.39  \\
\ion{Nd}{ii} && 2.54 &    1.03 && 1.99 &    0.59 && 2.23 &    0.48 && 1.93 &    0.38 && 2.07 &    0.56 && 2.02 &    0.37  \\
\ion{Sm}{ii} && 1.58 &    0.53 && 1.31 &    0.37 && 1.63 &    0.34 && 1.40 &    0.31 && 2.06 &    1.01 && 1.46 &    0.27  \\
\ion{Eu}{ii} && 0.83 &    0.22 && 0.55 &    0.05 && 1.00 &    0.15 && 0.73 &    0.08 && 0.70 &    0.09 && 0.83 &    0.08  \\
\noalign{\smallskip} \hline
\end{tabular}
\begin{tablenotes}
\item[$\ast$] The NLTE-corrected Ba abundances are listed here.
\end{tablenotes}
\end{threeparttable}
\end{table*}

The uncertainties on the abundance determination arise from two main sources:
stochastic errors (random errors on the spectral line measurements, uncertainties on
the $\log$\,$gf$) and systematic errors primarily due to uncertainties in
the determination of the stellar atmospheric parameters. For elements such as Na, Mg
Al, Si, Ca, Ti, Cr, Fe, Ni, Sr, Y, Zr, Ce, Nd and Sm, the uncertainty on
the EWs has been estimated to be about 3.5\,m\AA\,(see Fig.\,\ref{fig:ew}).
The magnitude of the random error on the abundance of an element X is
estimated as $\sigma_{ran} = \sigma/\sqrt{N-1}$, where $\sigma$ is the
root mean square (rms) around the mean X abundance, and N is the number of
lines of the X element (with $N \geq 2$).
For elements Sc, Mn, Ba, La and Eu, whose abundances were derived using the spectral synthesis method,
the abundance scatter from multiple lines of one element could represent the random
uncertainty \citep{Tautvaisiene2021}. If the abundance of an element
in a sample star is determined using only one line, we do not provide the
random error. Systematic uncertainties are dominated by the errors
in the choice of stellar parameters. They were estimated by varying $T_{\rm eff}$ by 100\,K,
$\log$\,$g$ by 0.1\,dex, [Fe/H] by 0.1\,dex and $\xi_{\rm t}$ by 0.2\,km\,s$^{-1}$.
The total uncertainty on the abundance of the element X, $\log\epsilon$(X), is computed
as the quadratic sum of the stochastic and systematic errors.
The results are presented in Table\,\ref{tab:uncert} for two sample stars, and it can
be found that the error is generally on the order of 0.15\,dex or less.

\begin{table*}
\centering \scriptsize \caption {The uncertainties of the abundance
analysis for the sample stars HD\,119650 and HD\,125079.} \label{tab:uncert}
\setlength\tabcolsep{2.0pt}
\begin{tabular}{lcccccccccccccccccc}
\hline \noalign{\smallskip}
&& \multicolumn{6}{c}{HD\,119650} && \multicolumn{6}{c}{HD\,125079}\\
\cline{3-8} \cline{10-15}\\
\multicolumn{1}{c}{Species} && $\sigma/\sqrt{N-1}$ & $\Delta {T_{\rm eff}}$ & $\Delta {\log g}$ & $\Delta {\rm [Fe/H]}$ & $\Delta {\xi_{\rm t}}$
& $\sigma_{\rm total}$ && $\sigma/\sqrt{N-1}$ & $\Delta {T_{\rm eff}}$ & $\Delta {\log g}$ & $\Delta {\rm [Fe/H]}$ & $\Delta {\xi_{\rm t}}$ &
$\sigma_{\rm total}$ \\
   &&  &  (+100 K) & (+0.1 dex) & (+0.1 dex) & (+0.2 km\,s$^{-1}$) &   &&   & (+100 K) & (+0.1 dex) & (+0.1 dex) & (+0.2 km\,s$^{-1}$) &  \\
\noalign{\smallskip} \hline \noalign{\smallskip}
\ion{Fe}{i}  && 0.01 & 0.04 & 0.01 & 0.01 & 0.09 & 0.10 && 0.01 &	0.08 &	0.01 & 0.08 & 0.00 & 0.11 \\
\ion{Fe}{ii} && 0.02 & 0.11 & 0.05 & 0.03 & 0.06 & 0.14 && 0.01 &	0.05 &	0.02 & 0.09 & 0.01 & 0.11 \\
\ion{Na}{i}  && 0.03 & 0.08 & 0.01 & 0.02 & 0.07 & 0.11 && 0.03 &	0.07 &	0.01 & 0.03 & 0.00 & 0.08 \\
\ion{Mg}{i}  &&  --  & 0.08 & 0.02 & 0.02 & 0.05 & 0.10 && 0.01 &	0.09 &	0.03 & 0.05 & 0.02 & 0.11 \\
\ion{Al}{i}  && 0.04 & 0.08 & 0.00 & 0.00 & 0.02 & 0.09 &&   -- &	0.06 &	0.00 & 0.00 & 0.00 & 0.06 \\
\ion{Si}{i}  && 0.01 & 0.04 & 0.02 & 0.02 & 0.04 & 0.06 && 0.01 &	0.03 &	0.01 & 0.02 & 0.01 & 0.04 \\
\ion{Ca}{i}  && 0.01 & 0.11 & 0.01 & 0.01 & 0.10 & 0.15 && 0.01 &	0.09 &	0.03 & 0.06 & 0.01 & 0.11 \\
\ion{Sc}{ii} && 0.04 & 0.05 & 0.05 & 0.04 & 0.06 & 0.11 && 0.04 &	0.04 &	0.06 & 0.09 & 0.09 & 0.15 \\
\ion{Ti}{i}  && 0.01 & 0.15 & 0.00 & 0.02 & 0.12 & 0.19 && 0.02 &	0.12 &	0.01 & 0.08 & 0.00 & 0.15 \\
\ion{Ti}{ii} && 0.02 & 0.01 & 0.04 & 0.04 & 0.09 & 0.11 && 0.01 &	0.01 &	0.03 & 0.07 & 0.03 & 0.08 \\
\ion{Cr}{i}  && 0.02 & 0.14 & 0.01 & 0.02 & 0.13 & 0.19 && 0.02 &	0.11 &	0.02 & 0.10 & 0.01 & 0.15 \\
\ion{Cr}{ii} && 0.03 & 0.07 & 0.05 & 0.03 & 0.07 & 0.12 && 0.02 &	0.04 &	0.04 & 0.07 & 0.02 & 0.10 \\
\ion{Mn}{i}  && 0.01 & 0.04 & 0.05 & 0.04 & 0.06 & 0.10 && 0.03 &	0.06 &	0.01 & 0.08 & 0.06 & 0.12 \\
\ion{Ni}{i}  && 0.01 & 0.03 & 0.02 & 0.02 & 0.10 & 0.11 && 0.01 &	0.08 &	0.00 & 0.06 & 0.01 & 0.10 \\
\ion{Sr}{i}  &&   -- &   -- &   -- &   -- &   -- &   -- &&   -- &	0.13 &	0.01 & 0.17 & 0.02 & 0.22 \\
\ion{Y}{i}   &&   -- & 0.08 & 0.02 & 0.06 & 0.04 & 0.11 &&   -- &	0.14 &	0.00 & 0.01 & 0.01 & 0.14 \\
\ion{Y}{ii}  && 0.02 & 0.00 & 0.04 & 0.03 & 0.10 & 0.11 && 0.02 &	0.01 &	0.03 & 0.10 & 0.04 & 0.11 \\
\ion{Zr}{i}  && 0.03 & 0.13 & 0.01 & 0.04 & 0.04 & 0.15 &&   -- &	0.15 &	0.00 & 0.01 & 0.01 & 0.15 \\
\ion{Zr}{ii} && 0.03 & 0.01 & 0.04 & 0.03 & 0.06 & 0.09 && 0.06 &	0.01 &	0.05 & 0.09 & 0.03 & 0.13 \\
\ion{Ba}{ii} && 0.05 & 0.02 & 0.01 & 0.04 & 0.11 & 0.13 && 0.06 &	0.05 &	0.05 & 0.06 & 0.07 & 0.13 \\
\ion{La}{ii} && 0.00 & 0.02 & 0.04 & 0.03 & 0.06 & 0.08 && 0.07 &	0.02 &	0.05 & 0.06 & 0.04 & 0.11 \\
\ion{Ce}{ii} && 0.03 & 0.02 & 0.04 & 0.04 & 0.09 & 0.11 && 0.04 &	0.02 &	0.03 & 0.12 & 0.03 & 0.13 \\
\ion{Nd}{ii} && 0.02 & 0.03 & 0.04 & 0.04 & 0.10 & 0.12 && 0.03 &	0.03 &	0.04 & 0.09 & 0.04 & 0.11 \\
\ion{Sm}{ii} && 0.03 & 0.03 & 0.04 & 0.04 & 0.06 & 0.09 && 0.04 &	0.02 &	0.04 & 0.05 & 0.03 & 0.08 \\
\ion{Eu}{ii} && 0.03 & 0.01 & 0.04 & 0.04 & 0.02 & 0.07 && 0.02 &	0.00 &	0.05 & 0.04 & 0.02 & 0.07 \\
\noalign {\smallskip} \hline
\end{tabular}
\end{table*}

\subsection{Comparisons of abundance results with previous literature}

To assess the reliability of the elemental abundances derived in this work, we compared
our results with those from the literature for the common stars. To minimize uncertainties
arising from different stellar atmospheric parameters adopted, we only compared the
elemental abundances between previous works and our results for stars with discrepancies
satisfying $\Delta T_{\rm eff} <$ 200\,K, $\Delta\log$\,$g$ $<$ 0.20\,dex, $\Delta$[Fe/H] $< 0.20$\,dex
and $\Delta \xi_{\rm t} < 0.40$\,km\,s$^{-1}$. The comparison results are listed in
Table\,\ref{tab:compare-abun}. Most of the elements exhibit mean deviations of $\lesssim 0.2$\,dex
between this work and previous studies. However, our Mg abundances for the three common stars are
noticeably lower compared to previous research. The primary reason for this difference may lie in the
usage of different spectral lines for determining Mg abundances. We primarily utilize the spectral lines
at 4571 and 5711\,\AA~to determine Mg abundances in our sample stars.
Notably, \citet{Luck2014} did not provide the atomic lines used for abundance determination.
\citet{Pereira2011} used eight lines to determine the Mg abundance of HD\,101079.
There is only one common line (5711\,\AA) between \citet{Pereira2011} and our study. For this
line, we adopted the excitation potential $\chi$ = 4.34\,ev and oscillator strength $\log$\,$gf$ = $-$1.75,
and derived the equivalent width of EW = 135\,m\AA, which closely match the values reported
in  \citet{Pereira2011} ($\chi$ = 4.34\,ev, $\log$\,$gf$ = $-$1.75 and EW = 132\,m\AA).
\citet{Karinkuzhi2014} utilized two lines to ascertain the Mg abundance of HD\,125079. Unfortunately, there is
no common line between \citet{Karinkuzhi2014} and our analysis.
There is also a significant deviation in the Sr abundance
($\sim$ 0.76\,dex) for HD\,125079 between this work and \citet{Karinkuzhi2014}. The stellar
atmosphere parameters for this star in our work are similar to those in \citet{Karinkuzhi2014}.
The Sr abundance is determined using only the \ion{Sr}{i} line at $\lambda$ = 4607\,\AA~in both works.
However, the corresponding $\log$\,$gf$ in our work ($\log$\,$gf$ = 0.28, \cite{Liu2021}) is 0.62
larger than that in \citet{Karinkuzhi2014} ($\log$\,$gf$ = $-$0.57). Therefore, the deviation in the
Sr abundance for HD\,125079 between the two works can be attributed to the different $\log$\,$gf$
values adopted.

\begin{table*}
\centering
\scriptsize
\caption{Comparisons of element abundances with literature values.} \label{tab:compare-abun}
\setlength\tabcolsep{1pt}
\begin{threeparttable}
\begin{tabular}{lrrrrrrrrrrrrc}
\hline \noalign{\smallskip}
Star name & [\ion{Na}{i}/Fe] & [\ion{Mg}{i}/Fe] & [\ion{Al}{i}/Fe] & [\ion{Si}{i}/Fe] &	[\ion{Ca}{i}/Fe] & [\ion{Sc}{ii}/Fe]
& [\ion{Ti}{i}/Fe] & [\ion{Ti}{ii}/Fe] & [\ion{Cr}{i}/Fe] & [\ion{Cr}{ii}/Fe] & [\ion{Mn}{i}/Fe] & [\ion{Ni}{i}/Fe] & References \\
\noalign{\smallskip} \hline \noalign{\smallskip}
HD\,77912  & 0.49 & $-$0.06 &    0.07 &  0.08   &  0.02   &  0.11   & $-$0.04 &  0.35   &  0.15   &  0.47   &$-$0.06  & $-$0.01 & 1 \\
           & 0.57 &    0.20 &    0.24 &  0.28   &   0.17  &  0.08   &    0.15 &   --    &  0.24   &  --     &  0.27   &  0.11   & 2 \\
HD\,101079 & 0.26 & $-$0.02 & $-$0.04 &  0.07   & $-$0.04 &  0.16   & $-$0.07 &  0.16   & $-$0.03 &  0.20   &  0.05   &  0.03   & 1 \\
           & 0.20 &  0.14   &  0.23   &  0.19   &  0.10   &    --   & $-$0.06 &    --   &  0.02   &    --   &    --   &  0.03   & 3 \\
HD\,125079 & 0.12 & 0.00 & $-$0.22 &  0.07   &  0.05   &  0.19   & $-$0.03 &  0.12   &  0.03   &  0.21   &  $-$0.03   & $-$0.01 & 1 \\
           & 0.34 &  0.05   &    --   &    --   &  0.03   &  0.02   &  0.00   &  0.45   & $-$0.06 &    --   & $-$0.22 &  0.07   & 4 \\
\noalign {\smallskip} \hline
\end{tabular}
\end{threeparttable}
\centering
\scriptsize
\setlength\tabcolsep{1pt}
\begin{threeparttable}
\begin{tabular}{lrrrrrrrrrrrc}
\hline \noalign{\smallskip}
Star name & [\ion{Sr}{i}/Fe] & [\ion{Y}{i}/Fe] & [\ion{Y}{ii}/Fe] & [\ion{Zr}{i}/Fe] & [\ion{Zr}{ii}/Fe] & [\ion{Ba}{ii}/Fe] &
[\ion{La}{ii}/Fe] & [\ion{Ce}{ii}/Fe] & [\ion{Nd}{ii}/Fe] & [\ion{Sm}{ii}/Fe] & [\ion{Eu}{ii}/Fe] & References \\
\noalign{\smallskip} \hline \noalign{\smallskip}
HD\,77912  &    -- &  0.27 &  0.46 & 0.19 & 0.52 & 0.56  & 0.25 & 0.29 & 0.49 & 0.51 & 0.15 & 1 \\
           &  0.47 &  0.36 &   --  & 0.14 &  --  &   --  & 0.43 & 0.81 & 0.47 & 0.31 & 0.24 & 2 \\
HD\,101079 &  0.14 &  0.36 &  0.71 & 0.32 & 0.59 & 0.48  & 0.44 & 0.53 & 0.70 & 0.36 & 0.03 & 1 \\
           &    -- &    -- &  0.63 &  --  & 0.44 &   --  & 0.43 & 0.28	& 0.51 &   -- &   -- & 3 \\
HD\,125079 &  0.83 &  0.89 &  0.97 & 0.90 & 1.09 & 1.15  & 1.03 & 1.06 & 1.02 & 0.61 & 0.34 & 1 \\
           &  1.59 &    -- &  1.05 &   -- &   -- & 1.06  &   -- & 0.93	& 1.16 & 0.56 &   -- & 4 \\
HD\,160538 &    -- &  0.25 &    -- & 0.16 &   -- & 0.25	 & 0.14 & 0.33 & 0.59 & 0.37 & 0.05 & 1 \\
           &  0.26 &  0.60 &    -- & 0.33 &   -- & 0.43  & 0.18 & 0.00	&   -- &   -- &   -- & 5 \\
\noalign {\smallskip} \hline
\end{tabular}
\begin{tablenotes}
\item References: 1. This work; 2. Average abundances provided by \citet{Luck2014}; 3. \citet{Pereira2011};
4. \citet{Karinkuzhi2014}; 5. \citet{Merle2016}.
\end{tablenotes}
\end{threeparttable}
\end{table*}

\section{Discussion}

In this study, we conducted a chemical abundance analysis of 20 Ba-stars spanning metallicities
in the range of $-$0.86 to 0.38\,dex. A low metallicity star, HD\,130255
([Fe/H] = $-$0.86\,dex), is included in this work. The abundance analysis of these Ba-stars
with [Fe/H] in a large range can help us better studying the stellar formation and chemical
evolution of Ba-stars.

\subsection{The Ba classifications of HD 115927 and HD 160538}
\label{sect:class}

\begin{figure}
\centering
\includegraphics[width=\columnwidth]{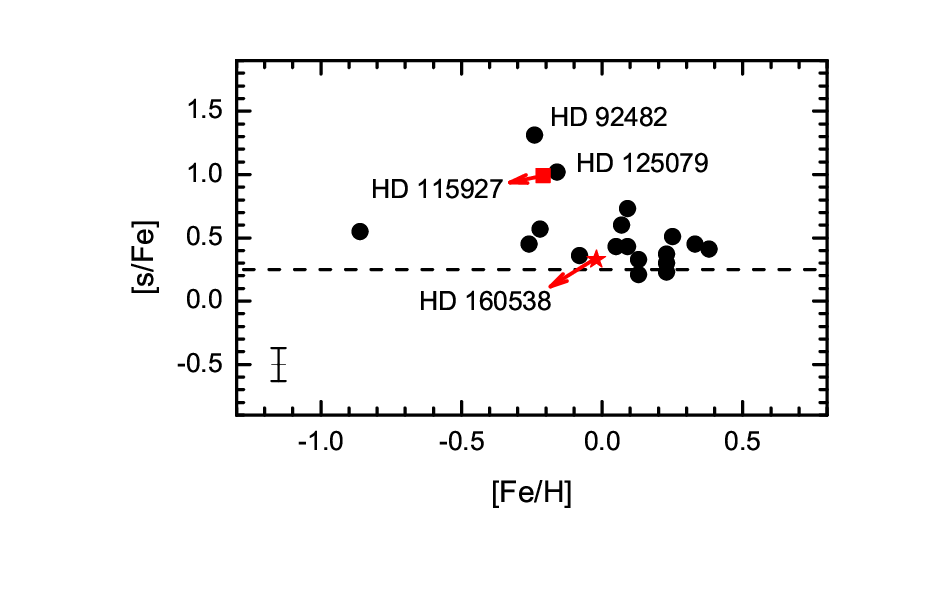}
\vspace{-1cm}
\caption{The ratios of [s/Fe] are plotted against [Fe/H] of the sample stars.
The red square and star refer to the two sample stars HD\,115927 and HD\,160538, respectively. The filled circles represent
the other 18 sample stars. The dashed line at [s/Fe] = 0.25\,dex represents the threshold for determining whether
a star is s-rich. The s-abundance of a star refers to the mean value of Sr, Y, Zr, Ba, La, Ce and Nd abundances.
The typical error bar is plotted in the bottom left-hand corner of the figure.}
\label{fig:feh-sfe}
\end{figure}

The [s/Fe] ratio serves as an important indicator for identifying whether a star is s-rich.
This ratio is defined as the mean abundance of the s-process elements, including Sr, Y, Zr, Ba,
La, Ce and Nd, relative to Fe abundance \cite[see for example Fig. 16 in][]{deCastro2016}. A star
is classified as s-rich if its [s/Fe] ratio is significantly higher than that in the other stars.
Various studies have proposed different thresholds for this classification. For instance,
\citet{Rojas2013} adopted [s/Fe] = 0.34\,dex, while \citet{deCastro2016} suggested 0.25\,dex.
In Fig.\,\ref{fig:feh-sfe}, we show [s/Fe] as a function of [Fe/H] for
the sample stars. The typical error is calculated based on the average [s/Fe] uncertainty
of the sample stars HD\,119650 and HD\,125079. It can be seen from the figure that the s-process
abundances of our sample stars fall within the range of 0.21 $\leq$ [s/Fe] $\leq$ 1.31\,dex.
The [s/Fe] ratios of HD\,107950 (0.23\,dex) and HD\,176524 (0.21\,dex) are slightly
lower than 0.25\,dex. However, taking into account the uncertainties on the abundances,
these two sample stars could be considered as mild (or marginal) Ba-stars.

In our sample, there are two stars, i.e., HD\,115927 and HD\,160538, having no Ba intensity
\citep{Warner1965,Lu1991}. \citet{Forsberg2019} derived high abundance ratios for HD\,115927
with [Zr/Fe] = 0.68\,dex, [La/Fe] = 1.00\,dex and [Ce/Fe] = 1.10\,dex. Similarly, based on the
abundances of Sr, Y, Zr, Ba, La and Ce reported by \citet{Merle2016}, the [s/Fe] of HD\,160538
is 0.30\,dex. However, these authors did not evidence the Ba-rich nature of HD\,115927
and HD\,160538. Accordingly, these two stars can be categorized as strong and mild Ba-stars, respectively.
We further determined the abundances of additional elements and discussed the Ba classifications for
these two stars by taking into account their [s/Fe]. As shown in Fig.\,\ref{fig:feh-sfe}, the [s/Fe]
ratio of the Ba-unknown star HD\,115927 ([s/Fe] = 0.99\,dex) is similar to that ([s/Fe] = 1.02\,dex)
of the strong Ba-star HD\,125079 \citep{Lu1991}. Thus, we classify HD\,115927 as
a strong Ba-star. For the Ba-likely star HD\,160538, its [s/Fe] ratio (= 0.33\,dex) is in the [s/Fe]
range of the mild Ba-stars \citep{Lu1991}. Consequently, we classify HD\,160538 as a mild Ba-star.
In summary, among our sample stars, HD\,92482, HD\,115927 and HD\,125079 belong
to strong Ba-stars, while the other 17 are mild Ba-stars. This enlarges the sample of Ba-stars,
including both strong and mild classifications.

\subsection{Abundances of Na to Ni}

\begin{figure}
\centering
\includegraphics[width=\columnwidth]{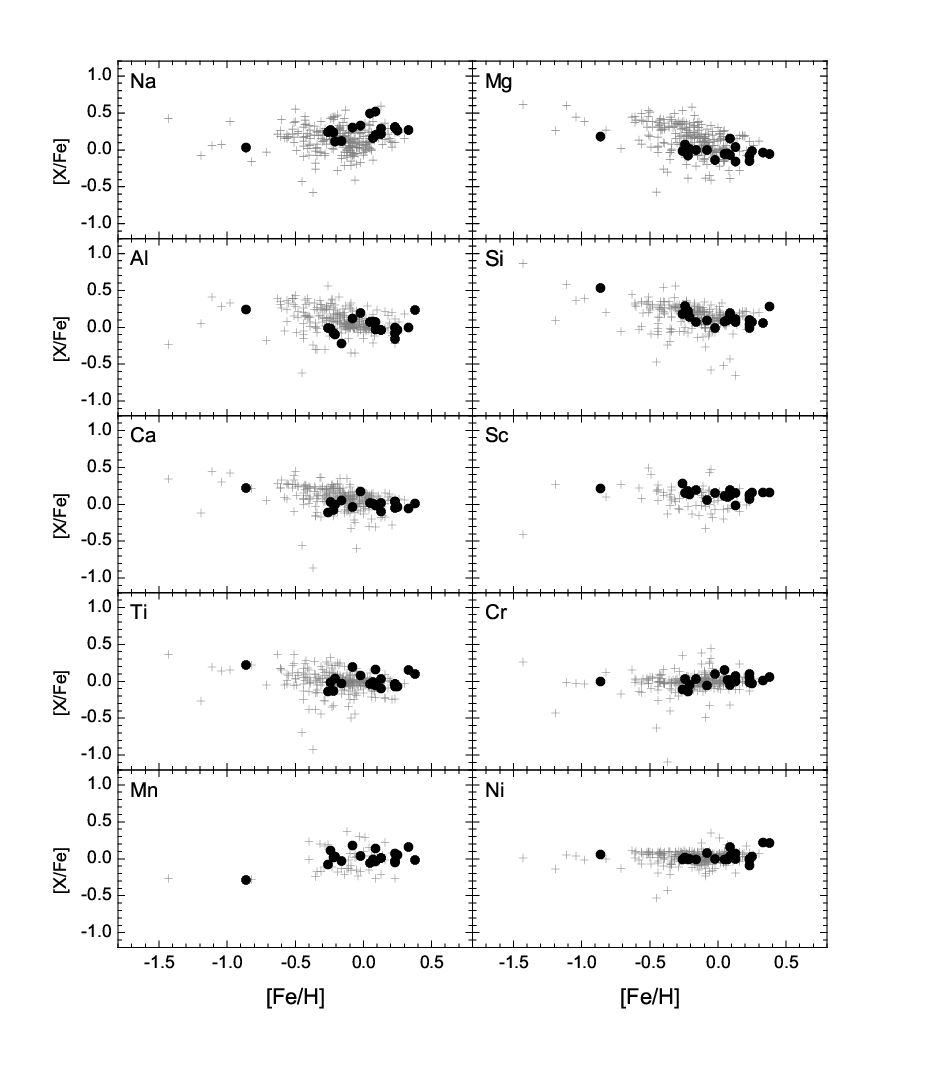}
\vspace{-0.5cm}
\caption{Abundance ratios [X/Fe] versus [Fe/H] for the elements from Na to Ni of the sample stars
in this work (filled circles) and those in previous studies (crosses):
\citet{Junqueira2001,Liang2003,Allen2006a,Smiljanic2007,Liu2009,Pereira2009,Pereira2011,deCastro2016,Mahanta2016,Yang2016}.}
\label{fig:feh-lfe}
\end{figure}

The chemical abundances of Ba-stars provide valuable insights into nucleosynthesis processes
across stars of different masses and the chemical evolution of binary or triple systems.
Fig.\,\ref{fig:feh-lfe} shows the abundance ratios [X/Fe] as a function of metallicity for
elements ranging from Na to Ni. Additionally, we included the abundances of 253 Ba-giants
analyzed by \citet{Junqueira2001,Liang2003,Allen2006a,Allen2006b,Smiljanic2007,Liu2009,Pereira2009,
Pereira2011,deCastro2016,Mahanta2016,Yang2016} for comparison. As shown in the figure, the abundance
distribution of most elements among our 20 Ba-stars (represented by filled circles) closely aligns
with those from previous studies (indicated by crosses).

As shown in Fig.\,\ref{fig:feh-lfe}, the abundance characteristics of these elements
in Ba-stars are similar to those in normal field stars. It is evident that the [Na/Fe] and [Al/Fe] ratios
exhibit no discernible trend with increasing [Fe/H]. The $\alpha$-elements Mg, Si, Ca and Ti are dominantly
generated in massive stars with M $> 10\,M_{\odot}$ \citep{Woosley1995}. In Fig.\,\ref{fig:feh-lfe},
the [X/Fe] ratios for all the $\alpha$-elements demonstrate a slight decrease with increasing [Fe/H].
In addition, the (normal) Mg abundances play in favor of low mass nature of the polluter AGB stars.
For the iron-peak elements Sc, Cr, Mn and Ni, although some scatter is observed, their abundances
remain consistent with those stars with corresponding metallicities. It is important
to note that the elemental abundances of neutral lines are adopted wherever available.

\subsection{Abundances of s-process elements}

\begin{figure*}
\centering
\includegraphics[width=15cm]{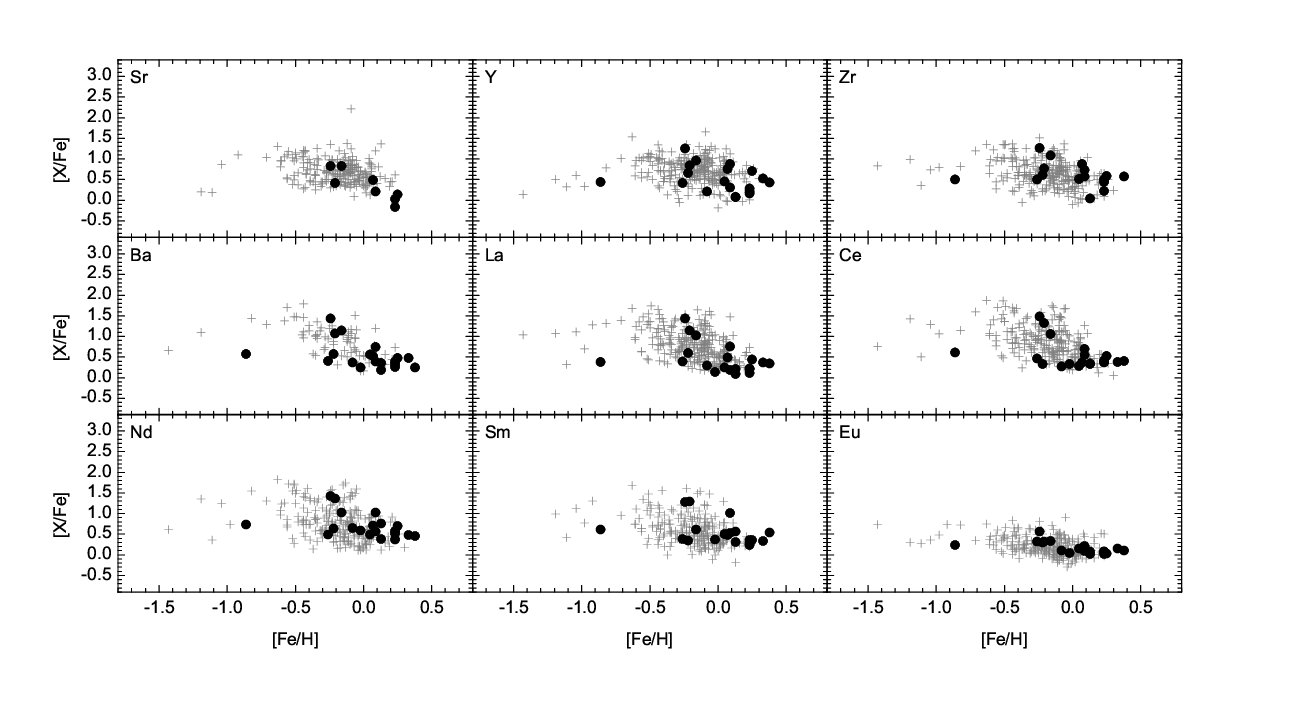}
\caption{Abundance ratios [X/Fe] versus [Fe/H] for the n-capture elements of Ba-stars.
The meanings represented by the filled circles and crosses are the same as in Fig.\,\ref{fig:feh-lfe},
but with the addition of data from \citet{Roriz2021b}.}
\label{fig:feh-nfe}
\end{figure*}

The chemical abundances of s-process elements serve as important tools in studying nucleosynthesis
processes and the formation history of Ba-stars. Fig.\,\ref{fig:feh-nfe} displays the [X/Fe] ratios
for the n-capture elements Sr, Y, Zr, Ba, La, Ce, Nd and Sm for the 20 sample stars (represented by
filled circles) in comparison with Ba-stars analyzed in previous studies (represented by crosses).
The abundance distribution of these s-process elements shows a good agreement with the previous
literature. From the figure, we can see that most of the Ba-stars exhibit overabundances of s-process
elements to varying degrees. It is worth noting that s-process nucleosynthesis contributes
about 30\% of the solar Sm \citep{Sneden2008}. Therefore, although its enhancements are appreciable
in most Ba-stars, as shown in Fig.\,\ref{fig:feh-nfe}, Sm is not considered as a truly s-process element.
Furthermore, the Galactic Chemical Evolution (GCE) models are not able to reproduce the abundance
levels of s-elements generally observed in Ba-stars \citep{Kobayashi2020}.

The lighter s-process elements Sr, Y and Zr constitute the first peak of
s-process abundances. Among our 20 sample stars, Sr abundances can be determined using \ion{Sr}{i}
line at 4607\,\AA~ for only eight stars, as no reliable abundance determinations are possible for any
\ion{Sr}{ii} lines in the spectra. Among these eight stars, seven exhibit slight enhancement of
Sr with 0.03 $\leq$ [\ion{Sr}{i}/Fe] $\leq$ 0.83\,dex, while the star HD\,107950 displays a mild
underabundance of Sr with [\ion{Sr}{i}/Fe] = $-$0.16\,dex. In LTE, the Sr abundances derived from
\ion{Sr}{i} line at 4607\,\AA~ are underestimated by about 0.1 to 0.2\,dex \citep{Bergemann2012}.
The discrepancy between LTE abundances derived from neutral and ionized lines can also
be seen for Y. It has been found that for stars close to solar metallicity,
the NLTE effects for all \ion{Y}{ii} lines are positive and do not exceed 0.12\,dex, while the
\ion{Y}{i} lines can reach up to $\sim$0.5\,dex \citep{Alexeeval2023}, therefore, we adopt the
Y abundance from \ion{Y}{ii} lines. It can be seen from Fig.\,\ref{fig:feh-nfe} that, Y and Zr abundances
are significantly overabundant relative to the solar values.

The heavier s-process elements Ba, La, Ce and Nd constitute the second peak of s-process abundances.
The abundances of heavier s-process elements in our Ba-stars exhibit a large scatter, and the patterns
in this work are consistent with those in previous literature. For our 20 sample stars, the abundances of
the typical s-process element Ba are measured from lines at 5853.67, 6141.71 and 6496.91\,\AA~.
The three strong Ba-stars HD\,92482, HD\,115927 and HD\,125079 exhibit strong enhancement of Ba with
[Ba/Fe] = 1.44, 1.07 and 1.15\,dex, respectively. The other 17 Ba-stars display mild enhancements
of Ba abundances with 0.19 $\leq$ [Ba/Fe] $\leq$ 0.75\,dex. As expected, all the sample stars exhibit
enhancement of not only Ba but also the other heavier s-process elements.

\begin{figure}
\centering
\includegraphics[width=\columnwidth]{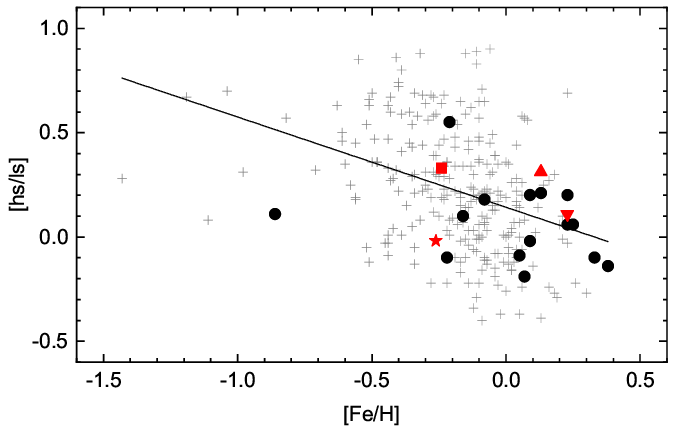}
\caption{Abundance ratios [hs/ls] versus [Fe/H] for Ba-stars. The red square, star, upper
and lower triangle represent the stars HD\,92482, HD\,150430, HD\,151101 and HD\,177304, respectively.
The filled circles and crosses are the same as in Fig.\,\ref{fig:feh-lfe}. The solid line is a linear
fit for the whole sample.}
\label{fig:hsls}
\end{figure}

The [hs/ls] ratios of Ba-stars serve as observational constraints on neutron exposures.
Consequently, these ratios have been widely used to indicate s-process efficiency and are frequently
incorporated by AGB nucleosynthesis models. Fig.\,\ref{fig:hsls} shows the [hs/ls] ratios as a function
of [Fe/H] for our sample stars, along with additional Ba-stars from the literature, similar to
Fig.\,\ref{fig:feh-lfe}. Here, [hs] and [ls] represent the mean abundances of heavier s-process elements
(Ba, La, Ce and Nd) and lighter s-process elements (Sr, Y and Zr), respectively. From the figure, we can
see that the [hs/ls] distribution of our sample stars aligns with the data in the literature.
The four stars HD\,92482, HD\,150430, HD\,151101 and HD\,177304, whose chemical abundances are analyzed for
the first time in the present work, have [hs/ls] ratios within the range of the literature, indicating that
these four objects are typical Ba-stars. The [hs/ls] ratios exhibit an anticorrelation with metallicity,
as discussed previously by \citet{Cseh2018} and \citet{Roriz2021b}, which indicates that at lower metallicities ([Fe/H] $<$ $-$0.6), the hs are the dominant products of
n-capture processes in AGB models. In contrast, at higher metallicities, the ls elements become predominant
\citep{Busso2001}. Negative values for [hs/ls] suggest that ls elements are more abundantly produced at the
corresponding metallicities.

\subsection{R-process characteristics of Ba-stars}

The typical r-process element Eu in our 20 sample stars exhibits slight overabundances with
0.02 $\leq$ [Eu/Fe] $\leq$ 0.57\,dex. For all the Ba-stars, the abundance scatter of Eu is obviously
smaller than that of both the lighter and the heavier s-process elements, which suggests that Eu has
not been significantly affected by the AGB ejecta, as depicted in Fig.\,\ref{fig:feh-nfe}.

Generally, Ba-stars display enhancements of s-process elements, with their abundance anomalies often
explained by binary or triple system scenarios \citep{Allen2006a,Stancliffe2021}. However, by comparing
the observed abundances of 12 Ba-stars with parametric model, \citet{Cui2014} found six Ba-stars having
significant r-process characteristics. They divided the Ba-stars into normal and r-rich Ba-stars.
\citet{Lugaro2012} proposed a stable multiple-star-system comprising a former AGB star and a former
SN\,II, to account for the stars with enrichment in both s- and r-process elements. The i-process,
initially suggested by \citet{Cowan1977}, is also a significant hypothesis for explaining the formation
of r/s stars \citep{Roederer2016,Goswami2020,Karinkuzhi2021}. However, \citet{Choplin2021}
found that the i-process occurring in AGB stars cannot fully explain the chemical abundances of giant stars
in binaries involving pollution by a former AGB companion.

\begin{figure}
\centering
\includegraphics[width=\columnwidth]{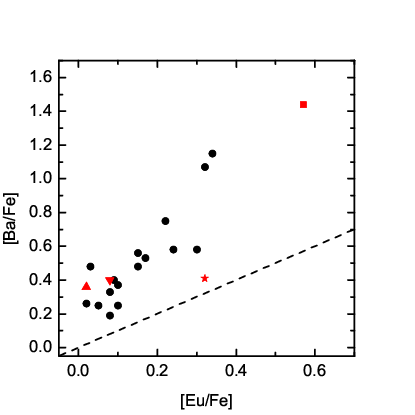}
\caption{Abundance ratios [Ba/Fe] versus [Eu/Fe] for the Ba-stars. The red square, star, upper
and lower triangle represent the stars HD\,92482, HD\,150430, HD\,151101 and HD\,177304, respectively.
The filled circles refer to the other 16 sample stars. The dashed line is a one-to-one correlation.}
\label{fig:bafe-eufe}
\end{figure}

Ba and Eu are considered as typical s-process and r-process elements, respectively, therefore many researchers
have analyzed the s- and r-process characteristics of stars based on the [Ba/Fe] and [Eu/Fe] ratios
\cite[e.g.,][]{Beers2005,Masseron2010}. Fig.\,\ref{fig:bafe-eufe} shows the abundance ratios [Ba/Fe] versus [Eu/Fe]
for our 20 Ba-stars. From the figure, we can see that all of the sample stars have [Ba/Eu] $>$ 0, indicating
that they all belong to normal Ba-stars. Among the four stars whose chemical abundances were determined for
the first time, HD\,92482 exhibits the highest [Ba/Fe] and [Ba/Eu] values, indicating its most pronounced
s-process characteristics. HD\,150430, HD\,151101 and HD\,177304 also lack r-process characteristics.

In addition to the approach that relies on two typical elements, \citet{Karinkuzhi2021} proposed
using the abundances of eight heavy elements to define a signed distance, $d_{s}$, for reclassifying
carbon-enhanced metal-poor (CEMP) stars as either CEMP-r, CEMP-s, or CEMP-rs.
Based on the new classification approach, they confirmed the CEMP-s stars that had been identified using
the abundance ratios of the typical s-process element La and the r-process element Eu. Additionally, they
reclassified two stars, previously classified as CEMP-rs, as CEMP-s objects.

Following the methodology outlined by \citet{Karinkuzhi2021}, we discuss the s- and r-characteristics of
the sample stars. This criterion is based on the signed distances ($d_{s}$) of a sample star from the solar
r-process abundance. The $d_{s}$ values are computed as follows:
\begin{equation}\label{dsequa}
d_{s}=\frac{1}{N}\sum_{i=1}^{N}(A_{i,\star}-A_{i,r,\odot})
\end{equation}
where, A$_{i,\star}$ = $\log_{10}(n_{i,\star}/n_{H,\star})$+12 with n$_{i,\star}$
the number density of a n-capture element i. A$_{i,r,\odot}$ is the solar r-process abundance
which is adopted from \citet{Goriely1999}, and has been scaled to the Eu
abundance of a target star. The element set i considered here is i = \{Sr, Y, Zr, Ba, La, Ce, Nd and Sm\}.
In this new classification scheme, $d_{s}$ increases for stars with higher s-process element enrichment.
This indicates that the observed abundance pattern deviates more significantly from the standard solar
r-process.
Following the definition described by \citet{Karinkuzhi2021}, $d_{s}$ = 0.6 is set as
separation between normal and r-rich Ba-stars. Table\,\ref{tab:distance} lists the derived $d_{s}$ values
of the 20 sample stars. As indicated in the table, all our sample stars are categorized as normal Ba-stars,
which means that the Ba enhancement observed in our sample stars should be attributed to material transfer
from the companions.

\citet{Karinkuzhi2021} applied this approach in the context of CEMP stars.
In their Fig.\,9, the CEMP-s stars have $d_{s}$ values in the range of about 0.6-1.4. As shown in
Table\,\ref{tab:distance}, the results for our Ba-stars are also in a similar range. This indicates
that the approach and threshold used for classifying CEMP stars may also be applicable in the identification
of Ba-stars. HD\,92482 has the highest $d_{s}$ value among all the stars in our sample, indicating that it
is the most s-rich star.

\begin{table}
\begin{center}
\caption[]{Signed distances of the sample stars.}
\label{tab:distance}
\begin{threeparttable}
\begin{tabular}{lclc}
  \hline\noalign{\smallskip}
Star name & $d_{s}$ & Star name & $d_{s}$ \\
  \hline\noalign{\smallskip}
HD\,77912  & 0.76 & HD\,133208 & 0.79 \\
HD\,85503  & 0.83 & HD\,136138 & 0.95 \\
HD\,92482  & 1.26 & HD\,150430 & 0.62 \\
HD\,101079 & 0.99 & HD\,151101 & 0.84 \\
HD\,107950 & 0.74 & HD\,160507 & 1.01 \\
HD\,115927 & 1.24 & HD\,160538 & 0.76 \\
HD\,119650 & 0.74 & HD\,168532 & 0.78 \\
HD\,125079 & 1.16 & HD\,176524 & 0.65 \\
HD\,130255 & 0.82 & HD\,176670 & 0.93 \\
HD\,131873 & 0.83 & HD\,177304 & 0.74 \\
\noalign{\smallskip}\hline
\end{tabular}
\end{threeparttable}
\end{center}
\end{table}

\subsection{Comparison of the observations to the models of AGB stars}

By using low-mass star models, \citet{Cristallo2011} investigated nucleosynthesis processes in AGB stars
and provided the FRUITY database dedicated to the nucleosynthesis occurring in AGB stars. The models span
a mass range from 1.3 to 6.0\,$M_{\odot}$ and metallicities $Z$ between 2 $\times$ 10$^{-5}$ and 3 $\times$ 10$^{-2}$.
The key assumptions of these models are: the primary neutron source is the radiative burning of $^{13}$C
within the pockets, and the efficiency of the third dredge-up (TDU) is inversely correlated with metallicity.
At a fixed metallicity, surface chemical enrichment of AGB stars mainly depends on the mass of dredged-up
material and the extent of the convective envelope.

In this study, we compared the n-capture abundance
patterns of our sample stars with the prediction of (non-rotation) FRUITY AGB models. It should be
noted that these models cannot be directly compared with the observations of Ba-stars because the effect of
dilution after mass transfer from the former AGB companions cannot be ignored. For the Ba-giants, the dilution
mainly results from the mass transfer and mixing. The mixing is mostly due to the first dredge-up, which mixes the
AGB material transferred from the companions ($M_{\rm AGB,trans}$) and the original envelope mass ($M_{\rm Ba,env}$).
In this case, the dilution factor is given by dil = ($M_{\rm AGB,trans}$+$M_{\rm Ba,env}$)/$M_{\rm AGB,trans}$
\citep{Becker1979,Bisterzo2012}.
The theoretical abundances of AGB models for individual Ba-stars were derived following the methodology outlined
in \citet{Cseh2022}. The diluted model abundances ([X/Fe]$_{\rm dil}$) were calculated as follows:
\begin{equation}\label{dilution}
\rm [X/Fe]_{dil}=\log(10^{[X/Fe]_{ini}}\times(1-\delta)+10^{[X/Fe]_{AGB}}\times\delta)
\end{equation}
where [X/Fe]$_{\rm ini}$ represents the initial abundances in the models, [X/Fe]$_{\rm AGB}$ indicates the final abundances
presented in the AGB models. The value of $\delta$ (= 1/dil) was derived by scaling the [Ba/Fe] of each AGB model
to that of the program star. Subsequently, the [X/Fe]$_{\rm dil}$ of each element of each model can be calculated by using
the same $\delta$ value. To identify the best-match model for a Ba-star, we preferentially selected models
with similar metallicity to the target star and then employed the $\chi^{2}$ method. We select models with
$\delta$ < 1, as $\delta$ = 1 is unrealistic since it implies that the AGB material is deposited on top of the Ba-star without
any mixing.

\begin{figure}
\centering
\includegraphics[width=\columnwidth]{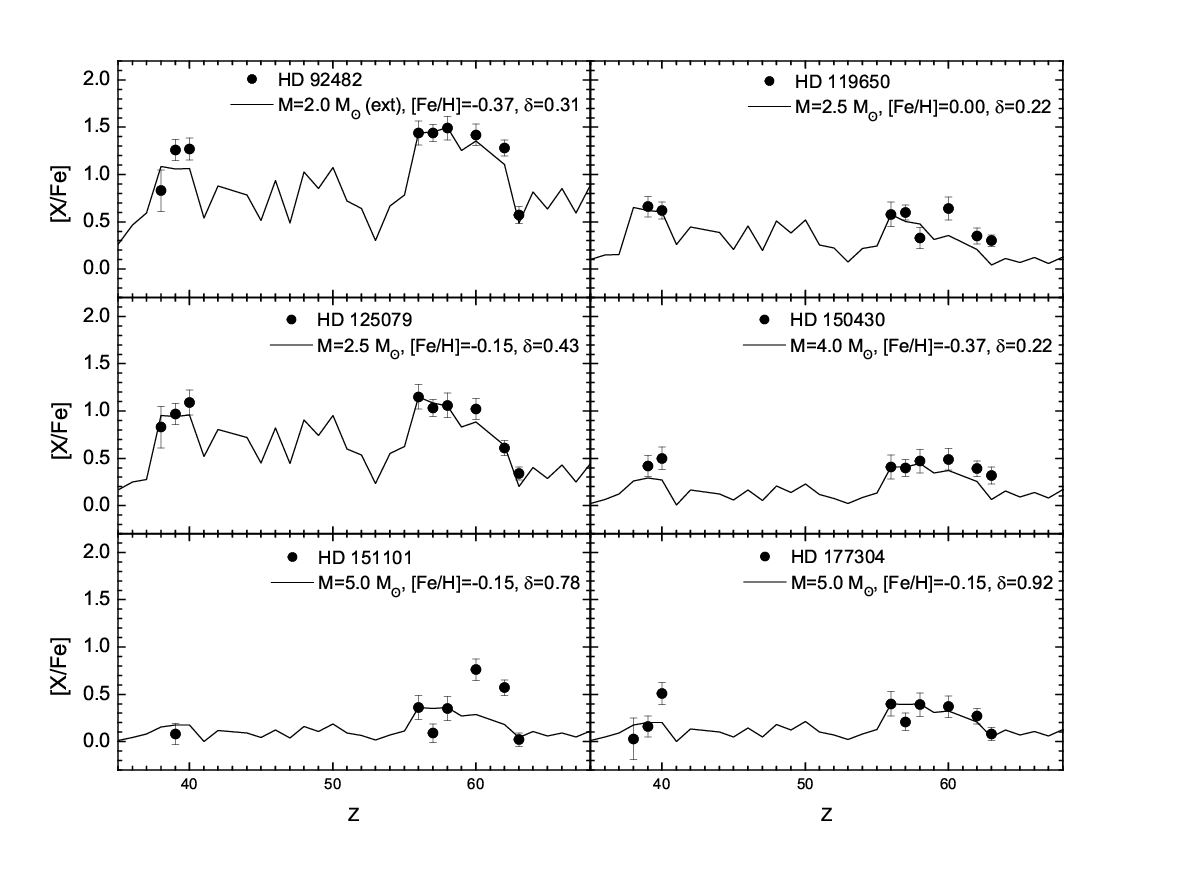}
\vspace{-1cm}
\caption{Comparison of the observed abundance patterns of the six sample stars
(represented by filled circles with error bars) with the FRUITY AGB models (shown as curves), which exhibit the closest
fits to the observational data following the application of dilution. ``Ext'' refers to the models with an extend $^{13}$C
pocket (labeled as `TAIL' in the FRUITY models).}
\label{fig:fruity}
\end{figure}

Taking the six stars, HD\,119650, HD\,125079, HD\,92482, HD\,150430, HD\,151101 and HD\,177304, as examples,
Fig.\,\ref{fig:fruity} shows the comparisons of abundance patterns between the target stars and the best-match models.
From the figure we can see that the diluted model patterns of AGB stars with $M$ = 2.0-5.0\,$M_{\odot}$ closely resemble
the observed abundance patterns of these six sample stars, suggesting that the n-capture elements of these Ba-stars are
accreted from the WD companions with an initial mass range from 2.0 to 5.0\,$M_{\odot}$.

Regarding the dilution factors we find, the calculated $\delta$ values are mostly below 0.3. However, some
systems require larger $\delta$ values, which is similar to the findings of \citet{Cseh2022}. We present the $\delta$ values
for selected sample stars in Fig.\,\ref{fig:fruity}. From the figure, it is clear that the $\delta$ values of models
with stellar masses $M \leq$ 4.0\,$M_{\odot}$ are around or below 0.30, whereas for models with higher stellar masses, e.g.,
$M =$ 5.0\,$M_{\odot}$, the $\delta$ values are much larger.

The estimated masses of the companions of all our sample stars are listed in Table\,\ref{tab:mass}. As shown in the table,
the initial masses of companions for the 17 mild Ba-stars range from 1.5 to 5.0\,$M_{\odot}$, with an average of approximately
3.8\,$M_{\odot}$. Meanwhile, the initial masses of companions for the 3 strong Ba-stars range from 2.0 to
3.0\,$M_{\odot}$, with an average value of 2.5\,$M_{\odot}$, which is slightly lower than that of the
companions of mild Ba-stars. Consequently, our results are consistent with numerous prior investigations demonstrating the
low-mass characteristic of TP-AGB stars, which have contaminated the envelopes of Ba-stars
\citep{Allen2006a,Cseh2018,Den2023,Goswami2023,Roriz2024}.

\begin{table}
\begin{center}
\caption[]{Mass estimations of companions of the Ba-stars.}
\label{tab:mass}
\begin{threeparttable}
\begin{tabular}{lclc}
  \hline\noalign{\smallskip}
Star name & Mass ($M_{\odot}$) & Star name & Mass ($M_{\odot}$) \\
  \hline\noalign{\smallskip}
HD\,77912  & 2.0 & HD\,133208 & 4.0 \\
HD\,85503  & 2.5 & HD\,136138 & 3.0 \\
HD\,92482  & 2.0 & HD\,150430 & 4.0 \\
HD\,101079 & 4.0 & HD\,151101 & 5.0 \\
HD\,107950 & 4.0 & HD\,160507 & 4.0 \\
HD\,115927 & 3.0 & HD\,160538 & 1.5 \\
HD\,119650 & 2.5 & HD\,168532 & 4.0 \\
HD\,125079 & 2.5 & HD\,176524 & 5.0 \\
HD\,130255 & 4.0 & HD\,176670 & 5.0 \\
HD\,131873 & 5.0 & HD\,177304 & 5.0 \\
\noalign{\smallskip}\hline
\end{tabular}
\end{threeparttable}
\end{center}
\end{table}

\begin{figure}
\centering
\includegraphics[width=\columnwidth]{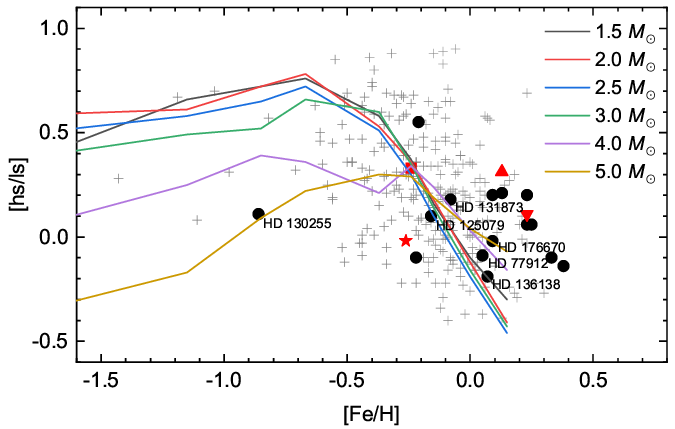}
\caption{[hs/ls] ratios as a function of [Fe/H] for the Ba-stars and the FRUITY AGB models with
the indicated stellar masses (shown as different colored curves). The symbols representing the Ba-stars are the
same as in Fig.\,\ref{fig:hsls}.}
\label{fig:fruity-hsls}
\end{figure}

In order to check the reliability of mass estimations by comparisons of the observed abundance patterns of Ba-stars
with the FRUITY models, we furthermore compared the [hs/ls] ratios of Ba-stars with those of FRUITY models with
different stellar masses, and presented the comparisons in Fig.\,\ref{fig:fruity-hsls}. From the figure, we can see
that models with stellar masses between 1.5-5.0\,$M_{\odot}$ can match the [hs/ls] ratios for [Fe/H] $\lesssim -$0.6.
However, for higher [Fe/H], these models do not cover the entire data range. In the figure, the data for HD\,130255,
HD\,131873 and HD\,176670 fall near the line of the 5.0\,$M_{\odot}$ model, aligning with the models selected in
Table\,\ref{tab:mass}. It is noteworthy that the abundance pattern of the 5.0\,$M_{\odot}$ ([Fe/H] = $-$0.85) model best
matches that of HD\,130255, however, it needs unexpected $\delta$ = 1.67 (> 1). Therefore, the mass of the companion star
is estimated to be 4.0\,$M_{\odot}$ with a reasonable $\delta$ of 0.50 (< 1), although the corresponding model-observation
match is slightly worse than that of the 5.0\,$M_{\odot}$ model. HD\,125079 lies on the line of the 2.5\,$M_{\odot}$ model.
HD\,77912 and HD\,136138 also fall within the region corresponding to models with stellar masses lower than 4.0\,$M_{\odot}$.
The distribution of [hs/ls] ratios of these six sample stars discussed above all align with the models we selected in
Table\,\ref{tab:mass}. However, the remaining sample stars either span a broad range within the FRUITY models or fall outside
their range, making the [hs/ls] ratios ineffective for selecting suitable AGB models for these stars.

In this section, we identified five FRUITY models with a stellar mass of 5.0\,$M_{\odot}$ that best fit
the observations. Typically, the masses of Ba-star companions are considered to be below 3.0 or 4.0\,$M_{\odot}$
\cite[e.g.,][]{Husti2009,Cseh2018,Shejeelammal2020,Roriz2021a}. However, \citet{Cseh2022} raised the question of where the
Ba-stars polluted by higher mass AGB companions are. They predicted that the stars with [s/Fe] < 0.25 are expected from
theoretical models of mass over 4.0\,$M_{\odot}$. We also notice that there are two sample stars (HD\,77912 and HD\,168532)
whose masses are higher than the companions we estimated. The masses of these two stars are estimated to be 4.5 and
6.0\,$M_{\odot}$ in this study. Generally, the typical masses of Ba-stars are found to be lower than 3.0\,$M_{\odot}$.
However, based on the orbital parameters and abundances of Ba-stars, \citet{Jorissen2019} predicted that the masses of
mild Ba-stars could extend up to 5.0\,$M_{\odot}$. Additionally, \citet{Roriz2021a} pointed out that, according to a
simple Salpeter initial mass function (IMF), there should be 14\% of Ba-stars with masses up to 4.0-6.0\,$M_{\odot}$.
These two unusual cases may reflect the complexity of mass transfer during the evolution of binary star systems.
To better understand these phenomena, further theoretical and observational efforts are required.

\section{Kinematic Analysis}

The analysis of star motions in the vicinity of the Sun relies on measurements of
their velocities along three key axes: $U$, $V$, and $W$. These axes correspond to directions
pointing towards the Galactic center, the path of Galactic rotation, and the North
Galactic Pole, respectively. By leveraging the precise stellar data of parallax, proper motion
and radial velocity, we can accurately calculate their velocities in space.
In our study, we acquired proper motion and parallax data for the 19 sample stars from
{\it Gaia} DR3 \citep{Gaia2023}. For HD\,131873 lacking proper motion and parallax measurements
in {\it Gaia} DR3, we acquired the data from SIMBAD database.

\begin{figure}
\centering
\includegraphics[width=\columnwidth]{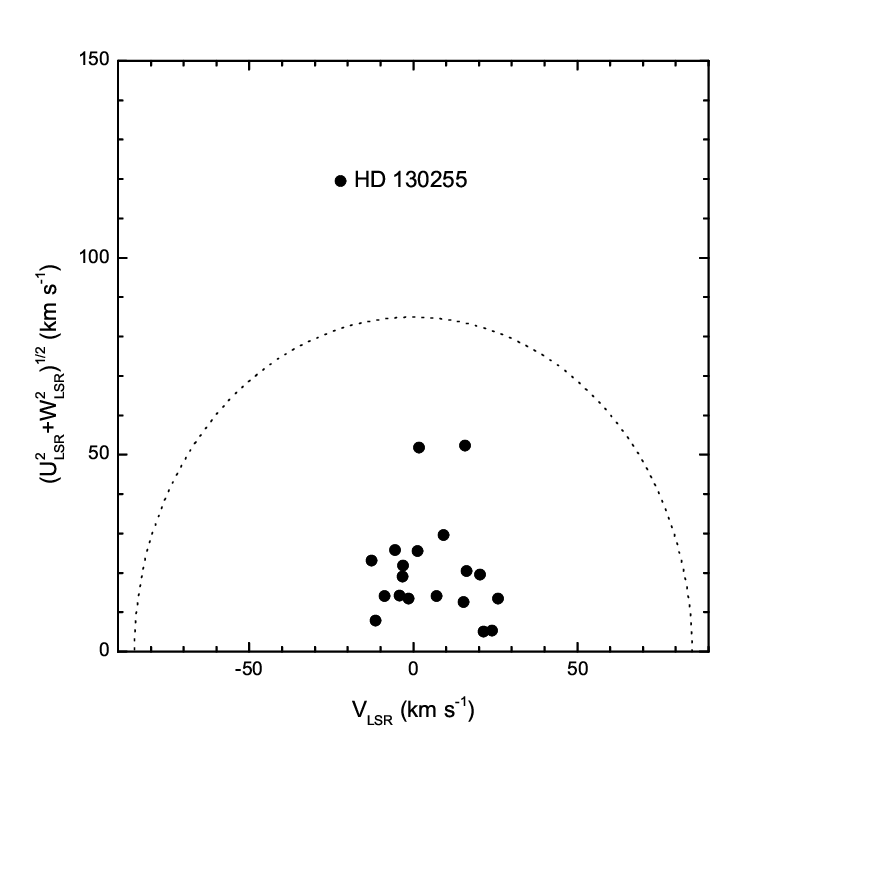}
\vspace{-2cm}
\caption{The Toomre diagram of ($U_{LSR}^2+W_{LSR}^2$)$^{1/2}$ versus $V_{LSR}$ for our sample stars.
The semi-circular line delineates the boundary for thin disk stars, with
$V_{\rm tot}$ = ($U_{LSR}^2 + V_{LSR}^2 + W_{LSR}^2$)$^{1/2}$ < 85 km\,s$^{-1}$ \citep{Chen2004}.}
\label{fig:UVW}
\end{figure}

The determination of the space velocity relative to the Sun follows the approach outlined by
\citet{Johnson1987}. Adjustments to the space velocity to align with the Local Standard of Rest
are made using a solar motion of ($U_{LSR}$, $V_{LSR}$, $W_{LSR}$)$_{\odot}$ = (11.1, 12.2, 7.3) km\,s$^{-1}$
\citep{Schonrich2010}. The Toomre diagram of ($U_{LSR}^2$+$W_{LSR}^2$)$^{1/2}$ as function of $V_{LSR}$
is presented in Fig.\,\ref{fig:UVW}. The semi-circular line delineates the boundary for thin disk stars, with
$V_{\rm tot}$ = ($U_{LSR}^2 + V_{LSR}^2 + W_{LSR}^2$)$^{1/2}$ < 85 km\,s$^{-1}$, as proposed by
\citet{Chen2004}.
From the figure, we can see that the majority of the stars in the sample are classified within the
thin disk, with the sole exception being the star HD\,130255, which can be classified as belonging
to the thick disk \cite[see also][]{deCastro2016}. HD\,130255 has the lowest metallicity ([Fe/H] = $-$0.86)
among our sample stars, which is consistent with the kinematic results.

Radial velocities of the 20 sample stars were determined through the Doppler shift of the elemental
absorption lines and were subsequently corrected for heliocentric motion. The derived results are
presented in Table \ref{tab:rv}. For comparison, values from {\it Gaia}\,DR3 and other literature sources
are also provided. All the sample stars are low-velocity objects ($V_{r}<$ 50 km\,s$^{-1}$). No literature
reported radial velocity variations for the sample stars. Our radial velocities for HD\,101079, HD\,115927,
HD\,119650, HD\,125079, HD\,136138, HD\,150430, HD\,160507, HD\,160538, HD\,168532 and HD\,176524 show
significant differences from the {\it Gaia}\,DR3 and literature values, which suggest that these stars
are in binary systems. For the remaining stars, we note only small differences from the literature values,
and those stars might actually be related to binaries with pole-on orbits or very eccentric orbits, where
significant radial velocity variations occur only in a small phase range \citep{Pourbaix2004,deCastro2016}.

\begin{table*}
\begin{center}
\caption[]{Comparisons of radial velocities with literature values.}
\label{tab:rv}
\begin{threeparttable}
\begin{tabular}{lrrrc}
  \hline\noalign{\smallskip}
Star name & $V_{r}$ (km\,s$^{-1}$) & $V_{r}$ (km\,s$^{-1}$) & $V_{r}$ (km\,s$^{-1}$) & References \\
          & (our estimates)        &         {\it Gaia}\,DR3         & (from the literature)   &            \\
  \hline\noalign{\smallskip}
 HD\,77912  &     15.9 &    16.1 & 15.4 & 1 \\
 HD\,85503  &     13.5 &    13.7 & 14.1 & 2\\
 HD\,92482  &   $-$7.2 &  $-$7.8 &  --  & --  \\
HD\,101079 &   $-$0.8 &  $-$3.9 &  $-$2.77 & 3 \\
 HD\,107950 &  $-$14.7 & $-$14.2 &  $-$14.2 & 1 \\
HD\,115927 &      1.4 &     4.2 &   2.7 & 4 \\
HD\,119650 &   $-$4.7 &  $-$8.3 & $-$5.7 & 5\\
HD\,125079 &   $-$4.0 &  $-$7.0 & $-$4.5 & 6 \\
 HD\,130255 &     41.5 &    41.0 & 40.6 & 7 \\
 HD\,131873 &     16.8 &    17.0 & 16.6 & 4 \\
 HD\,133208 &  $-$18.3 & $-$18.4 &  $-$18.2 & 1\\
HD\,136138 &   $-$5.9 & $-$10.8 &  $-$7.28 & 8 \\
HD\,150430 &      6.8 &     7.7 & 4.9 & 9 \\
 HD\,151101 &   $-$1.4 &  $-$1.4 & $-$1.0 & 10 \\
HD\,160507 &  $-$15.8 & $-$10.4 &  $-$14.5 & 11 \\
HD\,160538 &  $-$10.1 & $-$12.0 &  $-$9.0 & 10 \\
HD\,168532 &   $-$1.5 & $-$17.3 & $-$15.9 & 9 \\
HD\,176524 &  $-$16.5 & $-$10.0 &  $-$8.75 & 10 \\
 HD\,176670 &  $-$17.3 & $-$17.7 &  $-$17.7 & 9 \\
 HD\,177304 &  $-$10.6 & $-$11.3 &  --      & -- \\
  \noalign{\smallskip}\hline\\
\end{tabular}
\begin{tablenotes}
\item References: 1. \citet{Takeda2008}; 2. \citet{Hourihane2023}; 3. \citet{Pereira2011}; 4. \citet{Jonsson2017};
5. \citet{Gontcharov2006}; 6. \citet{Karinkuzhi2014} 7. \citet{deCastro2016}; 8. \citet{Massarotti2008}; 9. \citet{Famaey2005}; 10. \citet{Tautvaisiene2020};
11. \citet{Wang2011}.
\end{tablenotes}
\end{threeparttable}
\end{center}
\end{table*}

\section{Conclusions}

Based on the high resolution and high S/N spectra obtained from OHP, we derive the chemical abundances
of 20 elements (Na, Mg, Al, Si, Ca, Sc, Ti, Cr, Mn, Fe, Ni, Sr, Y, Zr, Ba, La, Ce, Nd, Sm and Eu) for
20 Ba-stars. The stellar atmospheric parameters of the sample stars cover a range of
4150 $\leq$$T_{\rm eff}$$\leq$ 5334\,K, 1.35 $\leq$ $\log$\,$g$ $\leq$ 3.20\,dex,
$-$0.86 $\leq$ [Fe/H] $\leq$ 0.38\,dex and 1.20 $\leq$ $\xi_{\rm t}$ $\leq$ 2.05\,km\,s$^{-1}$.
For the four sample stars, namely HD\,92482, HD\,150430, HD\,151101 and HD\,177304, it is the first time that
a detailed elemental abundances has been determined. Furthermore, among the remaining 16 sample stars,
Ba abundance has been derived for the first time in six of them. The conclusions are as follows:

\begin{enumerate}
  \item
  It is found that the [s/Fe] ratio of HD\,115927 is similar to that of the strong Ba-star HD\,125079, thus
  this star can be classified as a strong Ba-star. The [s/Fe] ratio of HD\,160538 is in the range of
  mild Ba-stars.
  \item
  The abundances of light odd-Z metal elements Na and Al, as well as Fe-peak elements Sc, Cr, Mn and Ni of
  Ba-stars remain similar to those of stars with corresponding metallcities. The [$\alpha$/Fe]
  ratios of Mg, Si, Ca and Ti show slight decrease with increasing [Fe/H].
  \item
  Most of the s-process elements in the Ba sample stars exhibit  varying degrees of overabundance. The first
  peak of s-process elements (Sr, Y and Zr) in most sample stars shows slight enrichment, while the second
  peak of s-process elements (Ba, La, Ce and Nd) in all Ba-stars shows significant overabundance.
  The [hs/ls] ratios increase with decreasing metallicity, which aligns with theoretical nucleosynthesis
  models of AGB stars using $^{13}$C as the neutron source. The abundance scatter of the r-process element
  Eu in Ba-stars is smaller than that of s-process elements, suggesting that very little Eu can be produced by
  the s-process in AGB stars.
  \item
  The signed distances for the sample stars indicate that all of them are normal Ba-stars, which means that
  the enhancements of s-process elements, such as Ba, La and Nd, should be attributed to material transfer
  from the companions.
  \item
  We derived the masses of the TP-AGB stars that previously polluted the Ba-stars. The observations
  can be effectively explained by the FRUITY nucleosynthesis models with low masses.
  \item
  Considering the kinematic perspective, we assessed the likelihood of these stars being part of the thin disk,
  thick disk, or halo. Our analysis revealed that the majority of these stars are associated with the thin disk,
  with HD\,130255 ([Fe/H] = $-$0.86) being the only one potentially affiliated with the thick disk.
\end{enumerate}

We anticipate that the derived chemical abundances of the 20 sample stars could contribute to a deeper
comprehension of the chemical evolution of Ba-stars. Further observational and theoretical studies involving
a larger sample of Ba-stars are imperative to explore the abundance characteristics of this stellar group more
comprehensively.

\section*{Acknowledgements}

We thank the anonymous referee for thoughtful suggestions that have improved the manuscript.
We appreciate Associate Professor Kefeng Tan from NAOC for the helpful discussions and constructive suggestions.
This work is supported by the National Natural Science Foundation of China (NSFC) under grant Nos. 11988101,
12273055, 11927804, 11673007, 11733006, 12173013, 12090040, and 12090044, the National Key R\&D Program of China
under grant No. 2019YFA0405502, the Fundamental Research Funds of China West Normal University under Grant No. 22kE041,
and the S\&T Program of Hebei under grant Nos. A2018106014 and A2021205006.
We acknowledge the support from the 2\,m Chinese Space Station Telescope project.

\section*{Data Availability}

This research is based on the spectral data retrieved from the ELODIE archive at Observatoire de Haute-Provence (OHP, http://atlas.obs-hp.fr/elodie/).




\begin{thebibliography}{99}
\bibitem[Abbott et al.(2017)]{Abbott2017} Abbott, B. P., Abbott, R., Abbott, T. D., et al., 2017, ApJL, 850, L39
\bibitem[Alexeeval et al.(2023)]{Alexeeval2023} Alexeeval, S., Wang, Y., \& Zhao, G., 2023, ApJ, 957, 10
\bibitem[Allen \& Barbuy(2006a)]{Allen2006a} Allen, D. M., \& Barbuy, B., 2006a, A\&A, 454, 895
\bibitem[Allen \& Barbuy(2006b)]{Allen2006b} Allen, D. M., \& Barbuy, B., 2006b, A\&A, 454, 917
\bibitem[Alonso et al.(1999)]{Alonso1999} Alonso, S., Arribas, S., \& Mart\'{i}nez-Roger, C., 1999, A\&AS, 140, 261
\bibitem[Antipova et al.(2004)]{Antipova2004} Antipova, L. I., Boyarchuk, A. A., Pakhomov, Y. V., \& Panchuk, V. E., 2004, Astronomy Reports, 48, 597
\bibitem[Asplund et al.(2009)]{Asplund2009} Asplund, M., Grevesse, N., Sauval, A. J., \& Scott, P., 2009, ARA\&A, 47, 481
\bibitem[Bard \& Kock(1994)]{Bard1994} Bard, A., \& Kock, M., 1994, A\&A, 282, 1014
\bibitem[Becker \& Iben(1979)]{Becker1979} Becker, S. A., \& Iben, I., Jr., 1979, ApJ, 232, 831
\bibitem[Beers \& Christlieb(2005)]{Beers2005} Beers, T. C., \& Christlieb, N., 2005, ARA\&A, 43, 531
\bibitem[Bensby et al.(2014)]{Bensby2014} Bensby, T., Feltzing, S., \& Oey, M. S., 2014, A\&A, 562, A71
\bibitem[Bergemann et al.(2012)]{Bergemann2012} Bergemann, M., Hansen, C. J., Bautista, M., \& Ruchti, G., 2012, A\&A, 546, A90
\bibitem[Bidelman \& Keenan(1951)]{Bidelman1951} Bidelman, W. P., \& Keenan, P. C., 1951, ApJ, 114, 473
\bibitem[Biemont et al.(1981)]{Biemont1981} Biemont, E., Grevesse, N., Hannaford, P., et al., 1981, ApJ, 248, 867
\bibitem[Bisterzo et al.(2012)]{Bisterzo2012} Bisterzo, S., Gallino, R., Straniero, O., et al., 2012, MNRAS, 422, 849
\bibitem[Boyarchuk et al.(2002)]{Boyarchuk2002} Boyarchuk, A. A., Pakhomov, Y. V., Antipova, L. I., \& Boyarchuk, M. E., 2002, Astronomy Reports, 46, 819
\bibitem[Burbidge et al.(1957)]{Burbidge1957} Burbidge, E. M., Burbidge, G. R., Fowler, W. A., \& Hoyle, F., 1957, RvMP, 29, 547
\bibitem[Busso et al.(2001)]{Busso2001} Busso, M., Lambert, D. L., Beglio, L., et al., 2001, ApJ, 557, 802
\bibitem[Casagrande et al.(2011)]{Casagrande2011} Casagrande, L., Sch\"{o}nrich, R., Asplund, M., et al., 2011, A\&A, 530, A138
\bibitem[Casamiquela et al.(2020)]{Casamiquela2020} Casamiquela, L., Tarricq, Y., \& Soubiran, C., 2020, A\&A, 635, A8
\bibitem[Chen et al.(2000)]{Chen2000} Chen, Y. Q., Nissen, P. E., Zhao, G., et al., 2000, A\&AS, 141, 491
\bibitem[Chen et al.(2004)]{Chen2004} Chen, Y. Q., Nissen, P. E., \& Zhao, G., 2004, A\&A, 425, 697
\bibitem[Choplin et al.(2021)]{Choplin2021} Choplin, A., Siess, L., \& Goriely, S., 2021, A\&A, 648, A119
\bibitem[Cowan \& Rose(1977)]{Cowan1977} Cowan, J. J., \& Rose, W. K., 1977, ApJ, 212, 149
\bibitem[Cowan et al.(2021)]{Cowan2021} Cowan, J. J., Sneden, C., Lawler, J. E., et al., 2021, RvMP, 93, 015002
\bibitem[Cristallo et al.(2009)]{Cristallo2009} Cristallo, S., Straniero, O., Gallino, R., et al., 2009, ApJ, 696, 797
\bibitem[Cristallo et al.(2011)]{Cristallo2011} Cristallo, S., Piersanti, L., Straniero, O., et al., 2011, ApJS, 197, 17
\bibitem[Cristallo et al.(2015)]{Cristallo2015} Cristallo, S., Straniero, O., Piersanti, L., \& Gobrecht, D., 2015, ApJS, 219, 40
\bibitem[Cristallo et al.(2016)]{Cristallo2016} Cristallo, S., Karinkuzhi, D., Goswami, A., et al., 2016, ApJ, 833, 181
\bibitem[Cseh et al.(2018)]{Cseh2018} Cseh, B., Lugaro, M., D'Orazi, V., et al., 2018, A\&A, 620, A146
\bibitem[Cseh et al.(2022)]{Cseh2022} Cseh, B., Vil\'{a}gos, B., Roriz, M. P., et al., 2022, A\&A, 660, A128
\bibitem[Cui et al.(2014)]{Cui2014} Cui, W. Y., Zhang, B., Shi, J. R., et al., 2014, A\&A 566, A16
\bibitem[de Castro et al.(2016)]{deCastro2016} de Castro, D. B., Pereira, C. B., Roig, F., et al., 2016, MNRAS, 459, 4299
\bibitem[Den Hartog et al.(2011)]{Den2011} Den Hartog, E. A., Lawler, J. E., Sobeck, J. S., et al., 2011, ApJS, 194, 35
\bibitem[Den Hartogh et al.(2023)]{Den2023} Den Hartogh, J. W., Yag\"{u}e, L. A., Cseh, B., et al., 2023, A\&A, 672, A143
\bibitem[De Smedt et al.(2015)]{desmedt2015} De Smedt, K., 2015, Ph.D. Thesis, The chemical diversity of post-AGB stars in the Galaxy and the Magellanic Clouds, KU Leuven, Belgium
\bibitem[Escorza et al.(2019)]{Escorza2019} Escorza, A., Karinkuzhi, D., Jorissen, A., et al., 2019, A\&A, 626, A128
\bibitem[Famaey et al.(2005)]{Famaey2005} Famaey, B., Jorissen, A., Luri, X., et al., 2005, A\&A, 430, 165F
\bibitem[Fern\'{a}ndez-Villaca\~{n}as et al.(1990)]{Fernandez1990} Fern\'{a}ndez-Villaca\~{n}as, J. L., Rego, M., \& Cornide, M., 1990, AJ, 99, 6
\bibitem[Fishlock et al.(2014)]{Fishlock2014} Fishlock, C. K., Karakas, A. I., Lugaro, M., \& Yong, D., 2014, ApJ, 797, 44
\bibitem[Forsberg et al.(2019)]{Forsberg2019} Forsberg, R., J\"{o}nsson, H., Ryde, N., \& Matteucci, F., 2019, A\&A, 631, A113
\bibitem[Fuhr et al.(1988)]{Fuhr1988} Fuhr, J. R., Martin, G. A., \& Wiese, W. L., 1988, JPRCD, 17, Suppl.4
\bibitem[Gaia Collaboration et al.(2023)]{Gaia2023} Gaia Collaboration, Vallenari, A., Brown, A. G. A., et al., 2023, A\&A, 674, A1
\bibitem[Gontcharov(2006)]{Gontcharov2006} Gontcharov, G. A., 2006, Astron. Lett., 32, 759
\bibitem[Goriely(1999)]{Goriely1999} Goriely, S., 1999, A\&A, 342, 881
\bibitem[Goriely \& Mowlavi(2000)]{Goriely2000} Goriely, S., \& Mowlavi, N., 2000, A\&A, 362, 599
\bibitem[Goswami \& Goswami(2020)]{Goswami2020} Goswami, P. P., \& Goswami, P., 2020, JAPA, 41, 47
\bibitem[Goswami \& Goswami(2023)]{Goswami2023} Goswami, P. P., \& Goswami, A., 2023, AJ, 165, 154
\bibitem[Green et al.(2019)]{Green2019} Green, G. M., Schlafly, E., Zucker, C., et al., 2019, ApJ, 887, 93
\bibitem[Han et al.(1995)]{Han1995} Han, Z. W., Eggleton, P. P., Podsiadlowski, P., \& Tout, C. A., 1995, MNRAS, 277, 1443
\bibitem[Hannaford et al.(1982)]{Hannaford1982} Hannaford, P., Lowe, R. M., Grevesse, N., et al., 1982, ApJ, 261, 736
\bibitem[Hourihane et al.(2023)]{Hourihane2023} Hourihane, A., Francois, P., Worley, C. C., et al., 2023, A\&A, 676A, 129H
\bibitem[Husti et al.(2009)]{Husti2009} Husti, L., Gallino, R., Bisterzo, S., et al., 2009, PASA, 26, 176
\bibitem[Ivans et al.(2006)]{Ivans2006} Ivans, I. I., Simmerer, J., Sneden, C., et al., 2006, ApJ, 645, 613
\bibitem[Ji \& Frebel(2018)]{Ji2018} Ji, A. P., \& Frebel, A., 2018, ApJ, 856, 138
\bibitem[Johnson \& Soderblom(1987)]{Johnson1987} Johnson, D. R. H., \& Soderblom, D. R., 1987, AJ, 93, 864
\bibitem[J\"{o}nsson et al.(2017)]{Jonsson2017} J\"{o}nsson, H., Ryde, N., Nordlander, T., et al., 2017, A\&A, 598, A100
\bibitem[Jorissen et al.(1998)]{Jorissen1998} Jorissen, A., Van Eck, S., Mayor, M., \& Udry, S., 1998, A\&A, 332, 877
\bibitem[Jorissen et al.(2019)]{Jorissen2019} Jorissen A., Boffin H. M. J., Karinkuzhi D., et al., 2019, A\&A, 626, A127
\bibitem[Junqueira \& Pereira(2001)]{Junqueira2001} Junqueira, S., \& Pereira, C. B., 2001, AJ, 122, 360
\bibitem[Karakas \& Lugaro(2016)]{Karakas2016} Karakas, A. I., \& Lugaro, M., 2016, ApJ, 825, 26
\bibitem[Karakas et al.(2018)]{Karakas2018} Karakas, A. I., Lugaro, M., Carlos, M., et al., 2018, MNRAS, 477, 421
\bibitem[Karinkuzhi \& Goswami(2014)] {Karinkuzhi2014} Karinkuzhi, D., \& Goswami A., 2014, MNRAS, 400, 1095
\bibitem[Karinkuzhi et al.(2021)] {Karinkuzhi2021} Karinkuzhi, D., Van Eck, S., Goriely, S., et al., 2021, A\&A, 645, A61
\bibitem[Katime et al.(2013)]{Katime2013} Katime Santrich, O. J., Pereira, C. B., \& de Castro, D. B., 2013, AJ, 146, 39
\bibitem[Kobayashi et al.(2020)]{Kobayashi2020} Kobayashi, C., Karakas, A. I., \& Lugaro, M., 2020, ApJ, 900, 179
\bibitem[Kong et al.(2018a)]{Kong2018a} Kong, X. M., Kumar, Y. B., Zhao, G., et al., 2018a, MNRAS, 474, 2129
\bibitem[Kong et al.(2018b)]{Kong2018b} Kong, X. M., Zhao, G., Zhao, J. K., et al., 2018b, MNRAS, 476, 724
\bibitem[Kramida et al.(2023)]{Kramida2023} Kramida, A., Ralchenko, Yu., Reader, J., and NIST ASD Team (2023). NIST Atomic Spectra Database (ver. 5.11). National Institute of Standards and Technology, Gaithersburg, MD.
\bibitem[Kurucz(1993)]{Kurucz1993} Kurucz, R. L., 1993, CD-ROM, Vol. 13, Smithsonian Astrophysics Observatory, Cambridge
\bibitem[Lambert et al.(1996)]{Lambert1996} Lambert, D. L., Health, J. H., Lemke, M., et al., 1996, ApJS, 103, 183
\bibitem[Lawler et al.(2019)]{Lawler2019} Lawler, J. E., Hala, Sneden, C., et al., 2019, ApJS, 241, 21
\bibitem[Lawler et al.(2001)]{Lawler2001} Lawler, J. E., Wickliffe, M. E., den Hartog, E. A., \& Sneden, C., 2001, ApJ, 563, 1075
\bibitem[Liang et al.(2000)]{Liang2000} Liang, Y. C., Zhao, G., \& Zhang, B., 2000, A\&A, 363, 555
\bibitem[Liang et al.(2003)]{Liang2003} Liang, Y. C., Zhao, G., Chen, Y. Q., et al., 2003, A\&A, 397, 257
\bibitem[Liu et al.(2009)]{Liu2009} Liu, G. Q., Liang, Y. C., \& Deng, L. C., 2009, RAA, 4, 431
\bibitem[Liu et al.(2021)]{Liu2021} Liu, S., Wang, L., Shi J. R., et al., 2021, RAA, 21, 278
\bibitem[Lu(1991)]{Lu1991} Lu, P. K., 1991, AJ, 101, 2229
\bibitem[Luck(2014)]{Luck2014} Luck, R. E., 2014, AJ, 147, 137
\bibitem[Luck(2015)]{Luck2015} Luck, R. E., 2015, AJ, 150, 88
\bibitem[Lugaro et al.(2012)]{Lugaro2012} Lugaro, M., Karakas, A. I., Stancliffe, R. J., \& Rijs, C., 2012, ApJ, 747, 2
\bibitem[Mahanta et al.(2016)]{Mahanta2016} Mahanta, U., Karinkuzhi, D., Goswami, A., et al., 2016, MNRAS, 463, 1213
\bibitem[Mashonkina et al.(1999)]{Mashonkina1999} Mashonkina, L. I., Gehren, T., \& Bikmaev, I. F., 1999, A\&A 343, 519
\bibitem[Mashonkina \& Gehren(2000)]{Mashonkina2000} Mashonkina, L., \& Gehren, T., 2000, A\&A, 364, 249
\bibitem[Massarotti et al.(2008)]{Massarotti2008} Massarotti, A., Latham, D. W., Stefanik, R. P., et al., 2008, AJ, 135, 209M
\bibitem[Masseron et al.(2010)]{Masseron2010} Masseron, T., Johnson, J. A., Plez, B., et al., 2010, A\&A, 509, A93
\bibitem[McClure et al.(1980)]{McClure1980} McClure, R. D., Fletcher, J. M., \& Nemec, J. M., 1980, ApJ, 238, L35
\bibitem[McClure(1983)]{McClure1983} McClure, R. D., 1983, ApJ, 268, 264
\bibitem[McWilliam(1990)]{McWilliam1990} McWilliam, A., 1990, ApJS, 74, 1075
\bibitem[McWilliam(1998)]{McWilliam1998} McWilliam, A., 1998, AJ, 115, 1640
\bibitem[Merle et al.(2016)]{Merle2016} Merle, T., Jorissen, A., Van, E. S., et al., 2016, A\&A, 586, A151
\bibitem[Mishenina et al.(2006)]{Mishenina2006} Mishenina, T. V.,  Bienaym\'{e}, O., Gorbaneva, T. I., et al., 2006, A\&A, 456, 1109
\bibitem[Moultaka et al.(2004)]{Moultaka2004} Moultaka, J., Ilovaisky, S. A., Prugniel, P., \& Soubiran, C., 2004, PASP, 116, 6933
\bibitem[Nissen \& Schuster(1997)]{Nissen1997} Nissen, P. E., \& Schuster, W. J., 1997, A\&A, 326, 751
\bibitem[North et al.(2000)]{Norris2000} North, P., Jorissen, A., \& Mayor, M., 2000, IAU Symp., 177, 269
\bibitem[O'Brian et al.(1991)]{OBrian1991} O'Brian, T. R., Wickliffe, M. E., Lawler, J. E., et al., 1991, JOSAB, 8, 1185
\bibitem[Pereira \& Drake(2009)]{Pereira2009} Pereira, C. B., \& Drake, N. A., 2009, A\&A, 496, 791
\bibitem[Pereira et al.(2011)]{Pereira2011} Pereira, C. B., Sales Silva, J. V., Chavero, C., et al., 2011, A\&A, 533, A51
\bibitem[Pourbaix et al.(2004)]{Pourbaix2004} Pourbaix, D., Tokovinin, A. A., Batten, A. H., et al., 2004, A\&A, 424, 727
\bibitem[Qian \& Wasserburg(2007)]{Qian2007} Qian, Y. Z., \& Wasserburg, G. J., 2007, PhR, 442, 237
\bibitem[Reddy \& Lambert(2017)]{Reddy2017} Reddy, A. B. S., \& Lambert, D. L., 2017, ApJ, 845, 151
\bibitem[Reetz(1999)]{Reetz1999} Reetz, J. K., 1999, PhD thesis, Ludwig-Maximilians-Univ. Munich
\bibitem[Roederer et al.(2016)]{Roederer2016} Roederer, I. U., Karakas, A. I., Pignatari, M., \& Herwig, F., 2016, ApJ, 821, 37
\bibitem[Roederer et al.(2018)]{Roederer2018} Roederer, I. U., Sakari, C. M., Placco, V. M., et al., 2018, ApJ, 865, 129
\bibitem[Rojas et al.(2013)]{Rojas2013} Rojas, M., Drake, N. A., Pereira, C. B., et al., 2013, Astrophysics, 56, 57
\bibitem[Roriz et al.(2021a)]{Roriz2021a} Roriz, M. P., Lugaro, M., Pereira, C. B., et al., 2021a, MNRAS, 501, 5834
\bibitem[Roriz et al.(2021b)]{Roriz2021b} Roriz, M. P., Lugaro, M., Pereira, C. B., et al., 2021b, MNRAS, 507, 1956
\bibitem[Roriz et al.(2024)]{Roriz2024} Roriz, M. P., Holanda, N., et al., 2024, AJ, 167, 184
\bibitem[Sch\"{o}nrich et al.(2010)]{Schonrich2010} Sch\"{o}nrich, R., Binney, J., \& Dehnen, W., 2010, MNRAS, 403, 1829
\bibitem[Shejeelammal et al.(2020)]{Shejeelammal2020} Shejeelammal, J., Goswami, A., Goswami, P. P., et al., 2020, MNRAS, 492, 3708
\bibitem[Smiljanic et al.(2007)]{Smiljanic2007} Smiljanic, R., Porto de Mello, G. F., \& da Silva, L., 2007, A\&A, 468, 679
\bibitem[Sneden et al.(2008)]{Sneden2008} Sneden, C., Cowan, J. J., \& Gallino, R., 2008, ARA\&A, 46, 241
\bibitem[Stancliffe(2021)]{Stancliffe2021} Stancliffe, R. J., 2021, MNRAS, 505, 5554
\bibitem[Takeda et al.(2008)]{Takeda2008} Takeda, Y., Sato, B., \& Murata, D., 2008, PASJ, 60, 781
\bibitem[Tautvai\v{s}ien\.{e} et al.(2020)]{Tautvaisiene2020} Tautvai\v{s}ien\.{e}, G., Mikolaitis, S., Drazdauskas, A., et al., 2020, ApJS, 248, 19T
\bibitem[Tautvai\v{s}ien\.{e} et al.(2021)]{Tautvaisiene2021} Tautvai\v{s}ien\.{e}, G., V\'{a}zquez, C. V., Mikolaitis, \v{s}., et al. 2021, A\&A, 649, 126
\bibitem[Wang et al.(2011)]{Wang2011} Wang, L., Liu, Y., Zhao, G., \& Sato, B., 2011, PASJ, 63, 1035
\bibitem[Warner(1965)]{Warner1965} Warner, B., 1965, MNRAS, 129, 263
\bibitem[Watson et al.(2019)]{Watson2019} Watson, D., Hansen, C. J., Selsing, J., et al., 2019, Nature, 574, 497
\bibitem[Woosley \& Weaver(1995)]{Woosley1995} Woosley, S. E., \& Weaver, T. A., 1995, ApJS, 101, 181
\bibitem[Yang et al.(2016)]{Yang2016} Yang, G. C., Liang, Y. C., et al., 2016, RAA, 16, 19
\bibitem[Yi et al.(2003)]{Yi2003} Yi, S. K., Kim, Y. C., \& Demarque, P., 2003, ApJS, 144, 259
\bibitem[Yushchenko et al.(2004)]{Yushchenko2004} Yushchenko, A. V., Gopka, V. F., Kim, C., et al., 2004, A\&A, 413, 1105
\bibitem[Za\v{c}s(1994)]{Zacs1994} Za\v{c}s, L., 1994, A\&A, 283, 937
\bibitem[Za\v{c}s et al.(1997)]{Zacs1997} Za\v{c}s, L., Musaev, F. A., Bikmaev, I. F., et al., 1997, A\&AS, 122, 31
\bibitem[Zhao et al.(2000)]{Zhao2000} Zhao, G., Qiu, H. M., Chen, Y. Q., \& Li, Z. W., 2000, ApJS, 126, 461
\bibitem[Zhao et al.(2016)]{Zhao2016} Zhao, G., Mashonkina, L., Yan, H. L., et al., 2016, ApJ, 833, 225
\end{thebibliography}



\bsp    
\label{lastpage}
\end{document}